\documentclass[preprint,12pt,nopreprintline]{elsarticle}

\usepackage{amssymb}
\usepackage{amsmath}
\usepackage{multirow}
\usepackage{tabularx}
\usepackage{booktabs} 
\usepackage{float}
\usepackage{graphicx} 
\usepackage{enumitem}
\usepackage{tikz}
\usepackage{array,booktabs,makecell}
\usetikzlibrary{arrows.meta,positioning,decorations.pathreplacing}
\usepackage{xcolor}

\usepackage{algorithm}
\usepackage{algpseudocode}

\usepackage{bm}
\usepackage[T1]{fontenc}
\usepackage[utf8]{inputenc}
\usepackage{adjustbox}

\usepackage{caption}
\usepackage{indentfirst}
\usepackage{geometry}
\usepackage[hidelinks]{hyperref}
\hypersetup{pdfauthor={},pdfsubject={},pdfkeywords={},pdfcreator={}}

\begin{document}

\begin{frontmatter}

\title{Adaptive Dead--Zone Dual Sliding Mode Observer for Reliable Electrochemical Model--Based SOC Estimation\tnoteref{published-version}}
\tnotetext[published-version]{This article has been published in \textit{Applied Energy}, Volume 427, Part D (January 2027), Article 129043; please cite the final published version: \href{https://doi.org/10.1016/j.apenergy.2026.129043}{doi:10.1016/j.apenergy.2026.129043}.}

\author[inst1]{Guangdi Hu} \author[inst2]{Keyi Liao} \author[inst1]{Jian Ye\corref{cor1}}
\author[inst3,inst4,inst5]{Feng Guo\corref{cor2}}

\affiliation[inst1]{organization={School of Vehicles and Intelligent Transportation,Fuyao University of Science and Technology},            city={Fuzhou},
            postcode={350109}, 
            country={China}}
\affiliation[inst2]{organization={School of Mechanical Engineering, Southwest Jiaotong University},            city={Chengdu},
            postcode={610031}, 
            country={China}}
\affiliation[inst3]{organization={VITO},            addressline={Boeretang 200}, 
            city={Mol},
            postcode={2400}, 
            country={Belgium}}    
\affiliation[inst4]{organization={EnergyVille},            addressline={Thor Park 8310}, 
            city={Genk},
            postcode={3600}, 
            country={Belgium}}

\affiliation[inst5]{    organization={Institute for Materials Research (IUMAT), Hasselt University},    addressline={Martelarenlaan 42},    city={Hasselt},    postcode={3500},    country={Belgium}}
      
\cortext[cor1]{Corresponding author:jianyee@fyust.org.cn(Jian Ye)}
\cortext[cor2]{Corresponding author:feng.guo@vito.be(Feng Guo)}

\begin{abstract}
Accurate state-of-charge (SOC) estimation is essential for the safe and efficient operation of lithium-ion batteries in electric vehicles and stationary energy storage systems. Electrochemical models enable high-fidelity SOC estimation, yet their practical deployment is challenged by nonlinear voltage dynamics, parameter drift, and the risk of divergence in joint state--parameter estimation. This paper proposes an adaptive dead-zone dual sliding-mode observer (ADZ-DSMO) based on an improved extended single-particle (IESP) model. In the proposed framework, a state SMO estimates SOC-related states, while a parameter SMO adapts dominant model parameters online. A Lyapunov-guided, online-computable adaptive dead-zone is developed to gate parameter updates, thereby suppressing error-propagation-induced divergence. Experimental validation on a Panasonic NCR18650PF cell under 1C, 3C, Urban Dynamometer Driving Schedule (UDDS), and New European Driving Cycle (NEDC) profiles shows that the proposed method achieves lower SOC error than the conventional dual sliding-mode observer (DSMO) and the IESP-based dual extended Kalman filter (DEKF). Under correct initialization, the SOC mean absolute error (MAE) remains below 0.25\% under 1C, UDDS, and NEDC, and reaches 1.13\% under the more demanding 3C condition. Under a 30\% initial SOC bias, the proposed method maintains MAE of 0.13--0.25\% under 1C, UDDS, and NEDC, and achieves 1.46\% MAE under 3C, with faster convergence than the DEKF in all tested cases. Moreover, selective parameter updating reduces the average computation time per sampling step compared with the always-updating DSMO, indicating improved real-time feasibility. Robustness is further evaluated under temperature variation and model-generated aging conditions.
\end{abstract}

\begin{keyword}
Lithium-ion battery \sep State-of-charge estimation \sep Electrochemical model \sep Dual sliding-mode observer \sep Lyapunov-guided adaptive dead-zone \sep Joint state--parameter estimation
\end{keyword}

\end{frontmatter}

\section{Introduction}
\label{sec1}

Lithium-ion batteries, owing to their high energy and power density compared with other rechargeable batteries, have become the dominant choice for pure electric vehicles, hybrid electric vehicles, as well as a variety of energy storage and power supply applications \cite{geng2025prospects,harper2019recycling,haghverdi2025review,liu2026adaptive,ding2023security}. Complex operating conditions require advanced battery management systems (BMSs) to ensure safe, reliable, and efficient operation. Accurate state-of-charge (SOC) estimation is a core BMS function that supports energy management, operating reliability, and battery longevity. To further enhance estimation accuracy, high-fidelity battery models are indispensable. In particular, electrochemical models, which capture the internal mechanisms of batteries, have attracted increasing attention from researchers for their potential in improving the reliability of SOC estimation \cite{guo2024systematic,wu2024physics, li2024applying}. In joint state--parameter estimation, transient state errors can drive biased parameter updates through the voltage residual. We address this coupling with an electrochemical-model-based adaptive dead-zone dual sliding-mode observer (ADZ-DSMO). Its novelty is an online-computable, Lyapunov-guided boundary that adapts to the estimated state and current input, rather than a fixed threshold. The boundary selectively regulates parameter updates while the state observer operates at every sampling step.

Battery models include ECMs~\cite{wang2025review}, electrochemical models~\cite{fuller1994simulation,guo2025control}, and data-driven models~\cite{GUO2026167,jing2025scalable,fan2025physics,li2025state,li2025physics,lin2025lightweight}. ECMs are efficient but rely on fitted parameters; data-driven methods require representative datasets and offer limited physical interpretability. Electrochemical models describe internal reaction and transport processes but are costly for onboard implementation. Reduced-order single-particle models retain key concentration dynamics with lower computational cost~\cite{li2019electrochemical,guo2025comparative,marquis2019asymptotic,haran1998determination}.
However, electrochemical-model-based SOC estimation is sensitive to parameter variations (aging, temperature, operating history) and model--plant mismatch, which motivates joint state--parameter estimation with online adaptation to maintain reliability over the battery life cycle.

Recent developments include physics-informed ECMs that represent solid-diffusion polarization for low-temperature SOC estimation~\cite{liu2026diffusion} and electrical--mechanical hybrid methods that incorporate additional information for SOC estimation~\cite{xiong2026electricalmechanical}. Electrochemically constrained fault-tolerant estimation also addresses voltage-sensor failures in flat-plateau LFP batteries~\cite{guo2026voltagesensors}.

Coulomb counting integrates current but accumulates drift~\cite{alzieu1997improvement}, whereas open-circuit-voltage methods require long relaxation periods~\cite{snihir2006battery,xing2014state}. Model-based observers use measured voltage to correct these errors in closed loop. The Kalman-filter (KF) family includes the extended Kalman filter (EKF)~\cite{plett2004extended,guo2019multi,peng2025state}, unscented Kalman filter (UKF)~\cite{zhao2025state,pang2025towards}, and cubature Kalman filter (CKF)~\cite{wang2025concurrent}. These methods support recursive state estimation, but their performance depends on model fidelity and noise assumptions. Particle filters (PFs) accommodate nonlinear and non-Gaussian dynamics, but the required particle population increases computational cost~\cite{lai2021novel,wang2020framework,chen2019particle}. In contrast, sliding mode observers (SMOs) \cite{qian2024switching, wei2025enhanced, feng2019robust, zhong2024nonlinear} offer several distinct advantages. SMOs provide robust correction against model uncertainties, disturbances, and parameter variations without requiring exact noise statistics. Moreover, SMOs are computationally efficient compared to PFs, making them well suited for real-time SOC estimation in practical BMSs. These advantages have motivated extensive research into improving SMO design, such as incorporating adaptive mechanisms and dead-zone strategies to mitigate chattering and enhance estimation reliability. 

Model parameters are commonly identified offline. Aging-related drift and calibration errors therefore motivate online adaptation for reliable SOC estimation. Joint estimation combines state correction with online parameter adaptation. Representative examples include recursive least square (RLS)-based methods \cite{zhang2018online}, dual Kalman filter algorithms \cite{guo2019equivalent,hu2012multiscale,shrivastava2021combined, chang2025state}, dual $H_\infty$ filters \cite{liu2025multi}, and dual particle filters \cite{xu2022online}. Most of these studies are based on ECMs. In recent years, however, several researchers have attempted to leverage more accurate electrochemical models for joint estimation of battery states and parameters. For instance, Gao et al. applied a dual extended Kalman filter (DEKF) to simultaneously estimate SOC and state of health (SOH) using an electrochemical model \cite{gao2021co}; Hosseininasab et al. employed a dual UKF for SOC estimation, achieving improved accuracy \cite{hosseininasab2023state}; Wang et al. introduced a dual particle filter based on electrochemical models for SOC estimation \cite{wang2024electrochemical}. Electrochemical models retain physically meaningful parameters that support SOC and SOH estimation, state-of-power (SOP) assessment, and fault diagnosis. Recent studies have further improved robust observer and control design under uncertainty. For example, Ding et al.~\cite{ding2023security} developed resilient event-triggered output-feedback lane-keeping control under denial-of-service attacks. In battery applications, Zhong et al.~\cite{zhong2024nonlinear} developed an adaptive super-twisting SMO within an electrochemical--thermal--aging model for battery microscopic-state estimation, and Zhong et al.~\cite{zhong2025safety} further used microscopic-state estimation to support safety-reinforced mutual pulse heating of lithium-ion battery packs. However, these methods mainly focus on robust state observation, safety-oriented control, or chattering attenuation. In contrast, electrochemical-model-based joint state--parameter estimation further requires reliable regulation of the parameter-update loop, because transient state-estimation errors or noisy voltage residuals may be propagated into parameter adaptation and lead to biased parameter correction.

Model-based estimation also supports structural isolation of concurrent electro-thermal faults in eVTOL battery systems~\cite{wang2026evtol} and sensor-enabled quantification of lithium plating under extreme environments~\cite{xiong2025plating}. Accelerated-aging studies further characterize capacity degradation under coupled thermal--electrical stresses~\cite{wang2026accelerated}.

However, simultaneous state and parameter estimation is prone to divergence, i.e., the algorithm may fail to converge to the correct values. This occurs because estimation errors in the state observer can propagate into the parameter estimator, resulting in biased parameter updates and eventual model deviation. Consequently, the application of joint state--parameter estimation has been limited. To address this issue, our previous work \cite{guo2018parameter} proposed a dead-zone strategy \cite{slotine1991applied} to improve the stability of joint estimation algorithms based on ECMs. Specifically, a fixed dead-zone dual extended Kalman filter (FDZ-DEKF) for tracking errors was developed and demonstrated effective mitigation of divergence issues under large initial state errors \cite{guo2018parameter}. The core idea was that by designing a parameter estimator with a dead zone, the transfer of state estimation errors into the parameter estimation loop could be reduced, thereby enhancing robustness. However, the dead zone in \cite{guo2018parameter} was defined as a fixed interval. Such a static design cannot completely block state estimation errors from entering the parameter estimation process. Over the entire battery life cycle, residual state errors can still accumulate, eventually causing divergence of the fixed dead-zone algorithm.

To address the above limitation, this paper develops a stability-guided joint state--parameter estimation framework based on a dual sliding-mode observer (DSMO) and an adaptive dead-zone. 
The contribution is a Lyapunov-guided, online-computable adaptive residual bound embedded in an electrochemical-model-based DSMO. It links parameter-update decisions to the evolving estimate, limiting the transfer of transient state errors into parameter adaptation.

The main contributions are summarized as follows:
\begin{itemize}
  \item \textbf{Electrochemical-model-based joint estimation with stability analysis:}
  We establish the coupled estimation-error dynamics for simultaneous SOC and parameter estimation in the proposed electrochemical model, and provide a discrete-time Lyapunov analysis that supports practical boundedness of the residual-driven state SMO and bounded conditional parameter adaptation under the proposed adaptive gate.

\item \textbf{Lyapunov-guided adaptive dead-zone for update gating:}
  Building on the general dead-zone idea, we construct an online-computable adaptive boundary using a nominal Lyapunov coefficient ratio to regulate parameter adaptation in the electrochemical-model-based DSMO.
  The boundary is computed using measurable current/voltage signals and estimated states, so that parameter updates are enabled only within a selected residual region and frozen when the residual is likely to be dominated by transient errors or uncertainties.
\item \textbf{Algorithmic transparency and validation under representative conditions:} A step-by-step computational procedure is provided for online implementation. The proposed method is evaluated under 1C and 3C constant-current conditions, UDDS and NEDC dynamic profiles, temperature variations, and model-generated aging scenarios.
\end{itemize}

Section~2 presents the model, battery tests, and parameter identification. Section~3 develops the observer and adaptive gate. Section~4 reports the results, and Section~5 concludes the study and outlines future work.

\section{An improved extended single-particle model}
The simplified electrochemical model used in this paper is an improved extended single
particle model (IESP). The IESP model is developed based on the extended single-particle model (ESPM)\cite{li2019electrochemical,li2018parameter}.

\subsection{Battery terminal voltage}

The input of the IESP model is the applied current $I$, and the output is the terminal voltage $U$. 
The terminal voltage is obtained by subtracting the concentration over-potential, reaction over-potential, 
and ohmic over-potential from the open-circuit voltage (OCV) \cite{li2019electrochemical,li2018parameter}:

\begin{equation}
U(t) = E_{\mathrm{OCV}}(t) - \eta_{\mathrm{con}}(t) - \eta_{\mathrm{act}}(t) - \eta_{\mathrm{ohm}}(t),
\label{eq:terminal_voltage}
\end{equation}

where  
\begin{itemize}
    \item $E_{\mathrm{OCV}}(t)$ is the open-circuit voltage determined by the electrode surface lithium-ion concentration,  
    \item $\eta_{\mathrm{con}}(t)$ is the concentration over-potential caused by the lithium-ion diffusion gradient,  
    \item $\eta_{\mathrm{act}}(t)$ is the activation over-potential associated with electrochemical reaction kinetics,  
    \item $\eta_{\mathrm{ohm}}(t)$ is the ohmic over-potential arising from the internal resistance of the electrolyte, electrodes, and current collectors.  
\end{itemize}

\subsection{Open-circuit voltage (OCV) and solid diffusion equations}

The OCV is directly determined by the lithium-ion concentration on the solid surface of 
the positive and negative electrodes. Assuming that the internal temperature of the battery 
remains constant, the OCV can be expressed as

\begin{equation}
E_{\mathrm{OCV}}(t) = U_{p}\!\left[x_{s,p,\mathrm{surf}}(t)\right] 
                 - U_{n}\!\left[x_{s,n,\mathrm{surf}}(t)\right],
\label{eq:ocv}
\end{equation}
where $U_p$ and $U_n$ are the equilibrium potentials of the positive and negative electrodes. The variables $x_{s,p,\mathrm{surf}}(t)$ and $x_{s,n,\mathrm{surf}}(t)$ are the corresponding normalized surface lithium concentrations (stoichiometries).  

The normalized solid-phase lithium concentration $x_{s,i}(r,t)$ inside a representative spherical particle evolves according to Fick's second law:

\begin{equation}
\frac{\partial x_{s,i}}{\partial t}(r,t)
=\frac{D_{s,i}}{r^{2}}
\frac{\partial}{\partial r}
\left(r^{2}\frac{\partial x_{s,i}}{\partial r}(r,t)\right),
\label{eq:solid_diffusion}
\end{equation}

where $i\in\{p,n\}$ denotes the positive ($p$) and negative ($n$) electrodes, $r\in[0,R_{s,i}]$ is the dimensional radial coordinate, $D_{s,i}$ is the solid-phase diffusion coefficient, and $R_{s,i}$ is the particle radius.  
A uniform initial concentration $x_{s,i}(r,0)=x_{i,0}$ is assumed.  

The boundary conditions accompanying~\eqref{eq:solid_diffusion} are
\begin{align}
\left.\frac{\partial x_{s,i}}{\partial r}\right|_{r=0}&=0, \label{eq:bc1}\\[4pt]
\left.\frac{\partial x_{s,i}}{\partial r}\right|_{r=R_{s,i}}&=-\frac{j_i(t)}{D_{s,i}}, \label{eq:bc2}
\end{align}
These conditions impose symmetry at the particle centre and the surface flux $j_i(t)$, expressed consistently with the normalized concentration $x_{s,i}$.  

Direct solution of Eq.~\eqref{eq:solid_diffusion} is computationally expensive. Therefore, spatial discretization or approximate methods are typically employed, including the parabolic approximation method, Padé approximation, spectral methods, and finite-difference methods \cite{guo2025comparative}.  
Among these, the parabolic approximation method \cite{luo2013approximate} is widely adopted because of its simplicity, computational efficiency, and relatively high accuracy.  

Using the parabolic approximation, the surface concentrations can be expressed as
\begin{equation}
x_{s,p,\mathrm{surf}}(t) = x_{s,p,\mathrm{avg}}(t) + \Delta x_{s,p}(t),
\label{eq:xspsurf}
\end{equation}

\begin{equation}
x_{s,n,\mathrm{surf}}(t) = x_{s,n,\mathrm{avg}}(t) + \Delta x_{s,n}(t),
\label{eq:xsnurf}
\end{equation}
where $x_{s,i,\mathrm{avg}}(t)$ is the average solid-phase concentration of electrode $i$ 
and $\Delta x_{s,i}(t)$ represents the deviation term describing the surface-to-average concentration difference.  

$Q_p$ and $Q_n$ represent the theoretical capacities of the effective active materials in the positive and negative electrodes, respectively \cite{li2019electrochemical}. 
They are defined as:
\begin{equation}
Q_p = \frac{Q_{\mathrm{all}}}{D_p},
\label{eq:Qp}
\end{equation}
\begin{equation}
Q_n = \frac{Q_{\mathrm{all}}}{D_n},
\label{eq:Qn}
\end{equation}
where $Q_{\mathrm{all}}$ denotes the theoretical maximum discharge capacity, 
and $D_n$ and $D_p$ are the maximum variation ranges of the stoichiometric numbers 
$x_{\mathrm{avg}}$ and $y_{\mathrm{avg}}$, respectively. 
Here, $x_{\mathrm{avg}}$ and $y_{\mathrm{avg}}$ are the normalized average lithium-ion  concentrations of the negative and positive electrodes.

The average solid concentrations are given by
\begin{equation}
x_{s,p,\mathrm{avg}}(t) = x_{s,p,0} + \int_{0}^{t} \frac{D_{p}}{Q_{\mathrm{all}}} I(\tau)\, d\tau,
\label{eq:xspavg}
\end{equation}

\begin{equation}
x_{s,n,\mathrm{avg}}(t) = x_{s,n,0} - \frac{D_{n}}{D_{p}} 
\left[ x_{s,p,\mathrm{avg}}(t) - x_{s,p,0} \right],
\label{eq:xsnavg}
\end{equation}

and the deviation terms follow first-order dynamics
\begin{equation}
\dot{\Delta x}_{s,p}(t) 
= \frac{1}{\tau_{s,p}}\left(\frac{12}{7}\frac{D_{p}\,r^{s}_{p}}{Q_{\mathrm{all}}} I(t) - \Delta x_{s,p}(t)\right),
\label{eq:dxsp_ct}
\end{equation}

\begin{equation}
\dot{\Delta x}_{s,n}(t) 
= \frac{1}{\tau_{s,n}}\left(\frac{12}{7}\frac{D_{n}\,r^{s}_{n}}{Q_{\mathrm{all}}} I(t) - \Delta x_{s,n}(t)\right).
\label{eq:dxsn_ct}
\end{equation}

By applying the forward Euler method with step size $\Delta t = t_{k+1}-t_k$, the discrete-time form is obtained:
\begin{equation}
\Delta x_{s,p}(t_{k+1}) = \Delta x_{s,p}(t_{k}) 
+ \frac{\Delta t}{\tau_{s,p}}
\left(\frac{12}{7}\frac{D_{p}\, r^{s}_{p}}{Q_{\mathrm{all}}} I(t_{k})
- \Delta x_{s,p}(t_{k})\right),
\label{eq:dxsp_disc}
\end{equation}

\begin{equation}
\Delta x_{s,n}(t_{k+1}) = \Delta x_{s,n}(t_{k}) 
+ \frac{\Delta t}{\tau_{s,n}}
\left(\frac{12}{7}\frac{D_{n}\, r^{s}_{n}}{Q_{\mathrm{all}}} I(t_{k})
- \Delta x_{s,n}(t_{k})\right).
\label{eq:dxsn_disc}
\end{equation}

In the ESP model, $\tau_{s,n}$ is usually regarded as a constant. 
However, in order to improve the model accuracy under high C-rate charge and discharge conditions, 
$\tau_{s,n}$ can be defined as a piecewise function of the current rate:
\begin{equation}
\tau_{s,n} =
\begin{cases}
\tau_{s,n,1}, & I < 1.5C, \\[6pt]
\tau_{s,n,2}, & 1.5C \leq I < 2.5C, \\[6pt]
\tau_{s,n,3}, & I \geq 2.5C.
\end{cases}
\label{eq:tausn_piecewise}
\end{equation}

The piecewise $\tau_{s,n}$ in Eq.~\eqref{eq:tausn_piecewise} is an effective (lumped) time constant in the reduced-order IESP under the parabolic approximation, introduced to capture rate-dependent voltage transients \cite{li2019electrochemical,li2018parameter}. In this work, $\tau_{s,n,1}$--$\tau_{s,n,3}$ are obtained by parameter identification via minimizing the terminal-voltage mismatch. Therefore, $\tau_{s,n}$ is not constrained to vary monotonically with $C$-rate and should not be interpreted as the intrinsic material diffusion time scale; rather, it reflects the best-fit relaxation speed of the $\Delta x_{s,n}$ state within each current-rate interval.

$Q_{\mathrm{all}}$ denotes the nominal battery capacity. In the ESP model, $Q_{\mathrm{all}}$ is assumed constant. 
In practice, however, the effective capacity decreases with cycling and with increasing charge/discharge current \cite{doerffel2006critical,deng2016online,omar2013peukert}.  
To account for this effect, Peukert's law is introduced to represent the variation of capacity under different $C$-rates \cite{omar2013peukert}.  
Although originally developed for lead--acid batteries, Peukert's law can be extended to LIBs.  The effective capacity is then given by \cite{yang2019comprehensive}:

\begin{equation}
Q_{\mathrm{now}} = Q_{\mathrm{all}} \left( \frac{C_{\mathrm{ref}}}{C_{\mathrm{now}}} \right)^{n-1},
\label{eq:peukert}
\end{equation}
where $C_{\mathrm{ref}}$ is the reference current rate (here $C_{\mathrm{ref}}=1C$), 
$C_{\mathrm{now}}$ is the applied current rate, 
$n$ is Peukert's constant, and $Q_{\mathrm{now}}$ is the effective capacity under the current operating condition. 
In subsequent formulations, $Q_{\mathrm{all}}$ is replaced by $Q_{\mathrm{now}}$. 
Peukert's law is employed here as a compact empirical correction to represent the rate--capacity effect observed in practical lithium-ion cells~\cite{yang2019comprehensive}. 
This correction does not explicitly alter the diffusion-state dynamics in Eqs.~\eqref{eq:solid_diffusion}--\eqref{eq:dc2_disc}; instead, $Q_{\mathrm{now}}$ enters the charge-balance/stoichiometry updates (e.g., Eqs.~\eqref{eq:xspavg}--\eqref{eq:xsnavg}), which shifts the resulting stoichiometries and surface concentrations $x_{s,i,\mathrm{surf}}$. 
Through Eqs.~\eqref{eq:ocv} and \eqref{eq:terminal_voltage}, this provides an indirect but implementation-friendly way to account for rate-dependent capacity variation in the reduced-order electrochemical model.

\subsection{Liquid diffusion}

The change of liquid-phase lithium-ion concentration 
at the positive and negative collectors is inconsistent. 
Thus, the liquid-phase diffusion over-potential can be expressed as
\begin{equation}
\eta_{\mathrm{con}}(t) = \frac{2RT}{F}(1-t_{+})
\ln \left( \frac{c_{0} + \Delta c_{1}(t)}{c_{0} - \Delta c_{2}(t)} \right),
\label{eq:eta_con}
\end{equation}
where $R$ is the universal gas constant, $T$ is the absolute temperature, 
$F$ is Faraday's constant, $t_{+}$ is the transference number, 
$c_{0}$ is the initial electrolyte concentration, and $\Delta c_{1}(t)$ and $\Delta c_{2}(t)$ 
are the concentration variations at the positive and negative electrodes, respectively.  

The discrete-time dynamics of $\Delta c_{1}$ and $\Delta c_{2}$ can be written as
\begin{equation}
\Delta c_{1}(t_{k+1}) = \Delta c_{1}(t_{k}) 
+ \frac{\Delta t}{\tau_{e}} \left[ P_{\mathrm{con},a}\, I(t_{k}) - \Delta c_{1}(t_{k}) \right],
\label{eq:dc1_disc}
\end{equation}

\begin{equation}
\Delta c_{2}(t_{k+1}) = \Delta c_{2}(t_{k}) 
+ \frac{\Delta t}{\tau_{e}} \left[ P_{\mathrm{con},b}\, I(t_{k}) - \Delta c_{2}(t_{k}) \right],
\label{eq:dc2_disc}
\end{equation}
where $\tau_{e}$ is the electrolyte diffusion time constant, 
$P_{\mathrm{con},a}$ and $P_{\mathrm{con},b}$ are empirical coefficients, 
and $\Delta t = t_{k+1}-t_{k}$ is the discrete sampling interval.

\subsection{Reaction polarization}

From the Butler--Volmer kinetic equation \cite{luo2013approximate}, the reaction polarization over-potential is obtained as
\begin{equation}
\eta_{\mathrm{act}}(t) = \frac{2RT}{F} 
\left\{ 
\ln \!\left[ \sqrt{m_{n}^{2}(t)+1} + m_{n}(t) \right] 
+ \ln \!\left[ \sqrt{m_{p}^{2}(t)+1} + m_{p}(t) \right] 
\right\},
\label{eq:eta_act}
\end{equation}
where $R$ is the universal gas constant, $T$ the absolute temperature, $F$ the Faraday constant, 
and $m_{p}(t)$ and $m_{n}(t)$ are intermediate variables related to electrode kinetics.  

The intermediate variables $m_{p}(t)$ and $m_{n}(t)$ can be expressed as

\begin{equation}
m_p(t)=
\frac{D_p P_{\mathrm{act}} I(t)}
{6Q_{\mathrm{now}}\sqrt{c_0}
\sqrt{x_{s,p,\mathrm{surf}}(t)
\left[1-x_{s,p,\mathrm{surf}}(t)\right]}},
\label{eq:mp}
\end{equation}

\begin{equation}
m_n(t)=
\frac{D_n P_{\mathrm{act}} I(t)}
{6Q_{\mathrm{now}}\sqrt{c_0}
\sqrt{x_{s,n,\mathrm{surf}}(t)
\left[1-x_{s,n,\mathrm{surf}}(t)\right]}},
\label{eq:mn}
\end{equation}

where $c_0$ is the initial electrolyte concentration, $P_{\mathrm{act}}$ is the reaction-polarization coefficient, and $0<x_{s,i,\mathrm{surf}}(t)<1$ is the surface stoichiometry.

\subsection{Ohm polarization}

Ohmic polarization follows Ohm's law, and the internal ohmic resistance of the battery 
can be obtained experimentally. Thus, the over-potential of ohmic polarization is given by
\begin{equation}
\eta_{\mathrm{ohm}}(t) = R_{\mathrm{ohm}} I(t),
\label{eq:eta_ohm1}
\end{equation}
where $R_{\mathrm{ohm}}$ is the internal ohmic resistance of the battery and $I(t)$ is the applied current.  

\subsection{Model parameter}

The battery used in the experiment is the Panasonic NCR18650PF cell. The experimental setup consists of an Arbin BT-5HC-5V/100A multifunctional battery tester, a Hongyu HY-TH-150DH temperature and humidity chamber,  an electronic computer for data acquisition and recording, and the NCR18650PF battery.

The cell specifications are summarized in Table~\ref{tab:battery_specs}.

\begin{table}[H]
\centering
\footnotesize
\caption{Specifications of the Panasonic NCR18650PF battery}
\begin{tabularx}{\linewidth}{lX}
\toprule
\textbf{Characteristic} & \textbf{Value} \\
\midrule
Rated capacity & Min. 2700 mAh; Typical 2900 mAh \\
Nominal voltage & 3.6 V \\
Voltage range & 2.5 V -- 4.2 V \\
Charge method & CC--CV, 4.2 V cutoff, 1365 mA standard current \\
Weight (max.) & 46 g \\
Energy density (volumetric / gravimetric) & 577 Wh/L; 207 Wh/kg \\
Positive electrode (cathode) material & LiNiMnCoO$_2$ (NMC) \\
Negative electrode (anode) material & Graphite \\
\bottomrule
\end{tabularx}
\label{tab:battery_specs}
\end{table}

The parameters were identified offline from experimental data following the procedure in 
\cite{li2019electrochemical}. Specifically, we used discharge datasets including 0.5C, 1C, 2C, 3C constant-current
profiles and the UDDS dynamic profile, and employed the Parameter Estimation toolbox in \textsc{Simulink} with the
Levenberg--Marquardt algorithm to minimize the terminal-voltage mismatch between measurements and simulations. 
The resulting fixed/estimated parameter values are summarized in Table~\ref{tab:params}.

\begin{table}[H]
\centering
\footnotesize
\caption{Fixed and estimated parameters of the improved extended single-particle model}
\begin{tabularx}{\linewidth}{l X c l}
\toprule
\textbf{Parameter} & \textbf{Description} & \textbf{Value} & \textbf{Unit} \\
\midrule
\multicolumn{4}{l}{\textit{Fixed parameters}} \\
\midrule
$F$ & Faraday constant & 96485.3 & C$\cdot$mol$^{-1}$ \\
$R$ & Ideal gas constant & 8.314 & J$\cdot$mol$^{-1}$$\cdot$K$^{-1}$ \\

$c_{0}$ & Initial electrolyte concentration & 1000 & mol$\cdot$m$^{-3}$ \\
\midrule
\multicolumn{4}{l}{\textit{Estimated parameters}} \\
\midrule
$D_{n}$ & Capacity-distribution coefficient (negative electrode) & 0.6533 & -- \\
$D_{p}$ & Capacity-distribution coefficient (positive electrode) & 0.7284 & -- \\
$P_{\mathrm{act}}$ & Reaction-polarization coefficient (BV auxiliary) & 90424 & s$\cdot$(mol$\cdot$m$^{-3}$)$^{1/2}$ \\
$P_{\mathrm{con},a}$ & Electrolyte diffusion gain (positive side) & 150 & mol$\cdot$m$^{-3}$$\cdot$A$^{-1}$ \\
$P_{\mathrm{con},b}$ & Electrolyte diffusion gain (negative side) & 60 & mol$\cdot$m$^{-3}$$\cdot$A$^{-1}$ \\

$R_{\mathrm{ohm}}$ & Ohmic internal resistance & 0.045 & $\Omega$ \\
$Q_{\mathrm{all}}$ & Total battery capacity & 2894.1 & mAh \\
$t_{+}$ & Li$^{+}$ transference number & 0.363 & -- \\
$\tau_{e}$ & Electrolyte diffusion time constant & 80 & s \\
$\tau_{s,n,1}$ & Solid diffusion time constant (negative) 1 & 1.1 & s \\
$\tau_{s,n,2}$ & Solid diffusion time constant (negative) 2 & 10 & s \\
$\tau_{s,n,3}$ & Solid diffusion time constant (negative) 3 & 0.05 & s \\
$\tau_{s,p}$ & Solid diffusion time constant (positive) & 1.85 & s \\
$x_{s,n,0}$ & Initial average solid stoichiometry (negative) & 0.745 & -- \\
$x_{s,p,0}$ & Initial average solid stoichiometry (positive) & 0.68 & -- \\
$n$ & Peukert exponent & 1.021 & -- \\
\bottomrule
\end{tabularx}
\label{tab:params}
\end{table}

\begin{figure}[H]
    \centering
    \includegraphics[width=\linewidth]{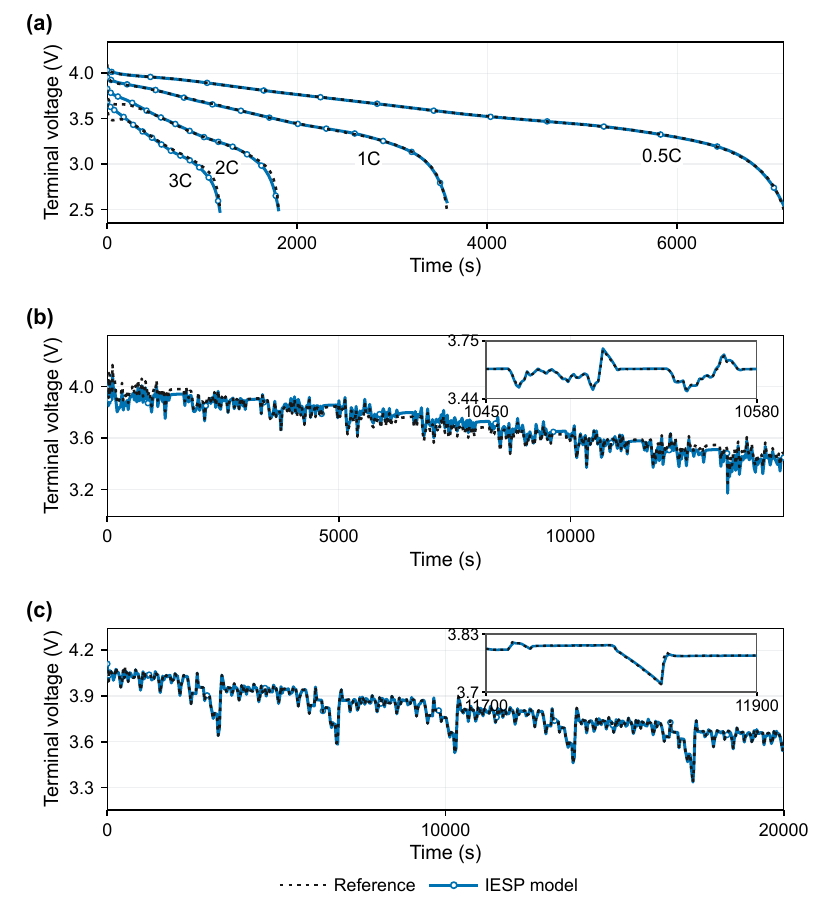}
    \caption{IESP terminal-voltage validation: (a) 0.5C, 1C, 2C, and 3C constant-current discharge; (b) UDDS; (c) NEDC. Insets resolve selected dynamic intervals.}
    \label{fig:model_simulation_result}
\end{figure}

The IESP model is implemented in MATLAB/Simulink. 
At room temperature (25~$^\circ$C), the input conditions include constant current discharges 
at 0.5C, 1C, 2C, and 3C rates, as well as the UDDS and NEDC. 
The UDDS and NEDC profiles represent typical dynamic charge/discharge scenarios of electric vehicles and are widely used to evaluate the ability of models to capture dynamic behavior.
The corresponding simulation results are shown in Figure~\ref{fig:model_simulation_result}.

As shown in Figure~\ref{fig:model_simulation_result}(a), the IESP model reproduces the overall terminal-voltage response of the battery under constant-current discharge conditions. 
At high C-rates (e.g., 2C and 3C), the measured voltage exhibits a pronounced initial drop, which is commonly attributed to strong polarization and an initially non-uniform lithium-ion concentration distribution within the electrodes, followed by a gradual recovery as the internal concentration field evolves. 
While a small mismatch may remain during this early-time transient, the IESP model remains consistent with the experimental voltage evolution over the full discharge duration and captures the main voltage variation trends across the considered C-rates. 
Under the dynamic UDDS and NEDC profiles (Figure~\ref{fig:model_simulation_result}(b,c)), the IESP model also tracks the measured terminal voltage well, with the voltage error remaining within about $\pm 0.05$~V. 
Overall, these results indicate that the developed IESP model provides a reliable basis for simulating terminal-voltage behavior under both static and dynamic operating conditions.

\section{Design of ADZ-DSMO Algorithm}
\label{sec:dual_smo}

This section presents the proposed ADZ-DSMO based on the IESP model established in Section~2. The observer consists of two coupled parts: a state SMO for estimating SOC-related states and a parameter SMO for updating the dominant electrochemical model parameters. Both observers use the terminal-voltage residual as the feedback signal, but they are updated differently: the state SMO is executed at every sampling step, whereas the parameter SMO is activated only when the residual satisfies a Lyapunov-guided adaptive bound. This design prevents transient state-estimation errors, measurement noise, or model mismatch from being directly propagated into the parameter adaptation loop. As a result, parameter updates are performed only within a stability-admissible region, which improves the reliability of long-term SOC estimation compared with continuous parameter updating or a fixed dead-zone strategy. Meanwhile, the selective-update mechanism reduces the number of parameter SMO executions by skipping unreliable or unnecessary parameter updates, thereby lowering computational burden and improving online computational efficiency.

\begin{figure}[H]
\centering
\makebox[\linewidth][c]{\includegraphics[width=1.22\linewidth]{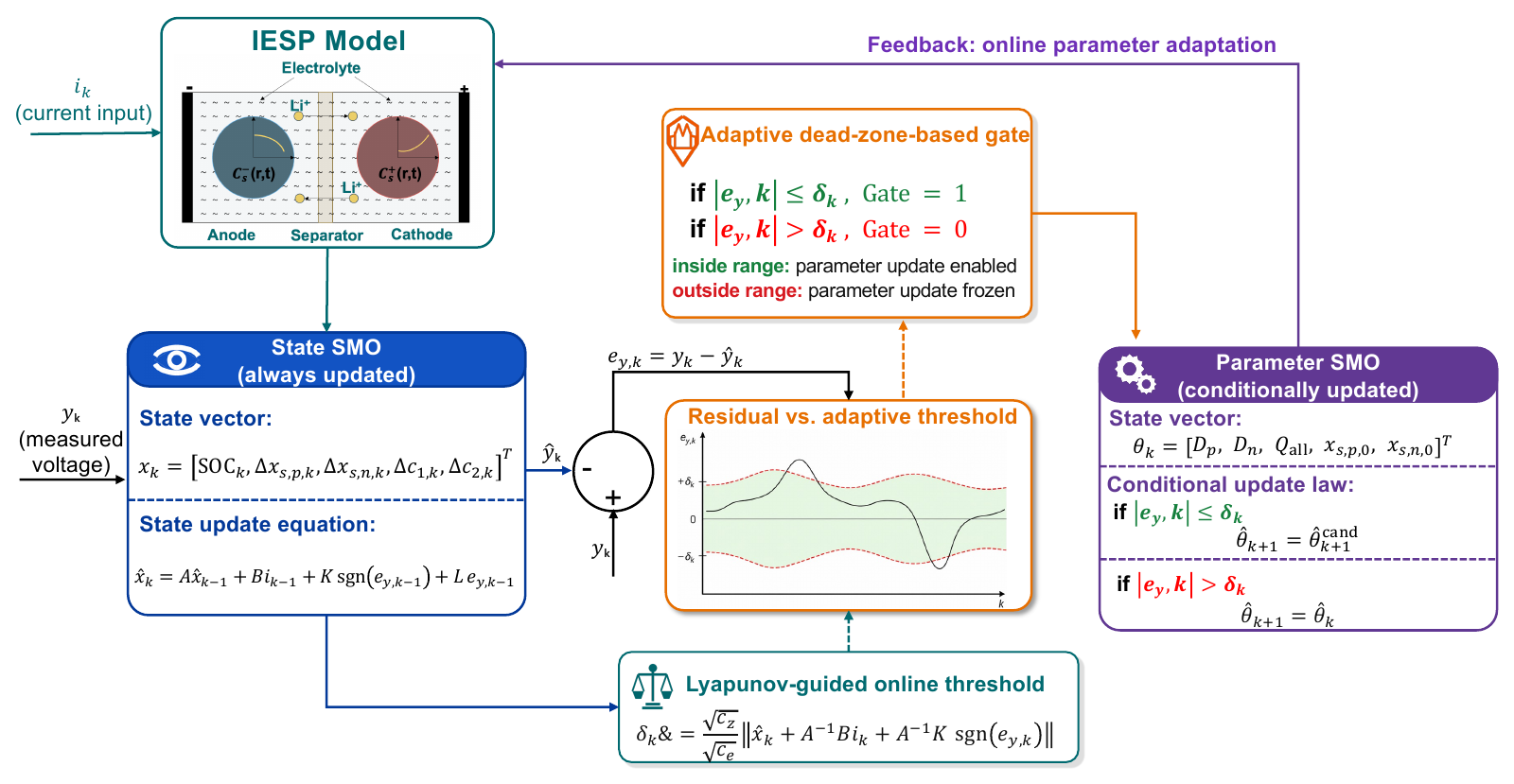}}
\caption{Schematic of the ADZ-DSMO. The state SMO runs at every sampling step; the gate conditionally enables the parameter SMO.}
\label{fig:adz_dual_smo_framework}
\end{figure}

The resulting workflow is illustrated in Figure~\ref{fig:adz_dual_smo_framework}. The measured current and terminal voltage are first used to update the state SMO and compute the estimated voltage. The residual \(e_{y,k}=y_k-\hat{y}_k\) is then compared with the adaptive threshold \(\delta_k\). If \(|e_{y,k}|\leq\delta_k\), the parameter SMO is updated; otherwise, parameter adaptation is frozen to avoid unreliable parameter correction. The following subsections define the state SMO, the parameter SMO, and the Lyapunov-guided adaptive dead-zone gate.

\subsection{State SMO Design}
\label{subsec:state_smo}

To be consistent with the IESP model in Section~2, we define the discrete-time state vector as

\begin{equation}
\bm{x}_k \triangleq 
\begin{bmatrix}
\mathrm{SOC}_k &
\Delta x_{s,p,k} &
\Delta x_{s,n,k} &
\Delta c_{1,k} &
\Delta c_{2,k}
\end{bmatrix}^{\top},
\label{eq:state_vector_def}
\end{equation}

where $\mathrm{SOC}_k$ is the state of charge, $\Delta x_{s,p,k}$ and $\Delta x_{s,n,k}$ denote the surface-to-average solid-phase concentration deviations defined in Eqs.~\eqref{eq:xspsurf}--\eqref{eq:xsnurf} and governed by Eqs.~\eqref{eq:dxsp_disc}--\eqref{eq:dxsn_disc}, and $\Delta c_{1,k}$ and $\Delta c_{2,k}$ are the electrolyte concentration-variation states governed by Eqs.~\eqref{eq:dc1_disc}--\eqref{eq:dc2_disc}.

The SOC state follows the Coulomb-counting relation
\begin{equation}
\mathrm{SOC}_{k} = \mathrm{SOC}_{k-1} - \frac{\Delta t}{Q_{\mathrm{now}}}\, I(t_{k-1}),
\label{eq:soc_cc}
\end{equation}
Here, $\Delta t$ is the sampling interval and $Q_{\mathrm{now}}$ is the effective capacity defined in Eq.~\eqref{eq:peukert}. Discharge current is positive. With current in amperes and time in seconds, capacity is expressed in coulombs ($1~\mathrm{mAh}=3.6~\mathrm{C}$).

The discrete-time state equation is

\begin{equation}
\bm{x}_{k} = \bm{A} \bm{x}_{k-1} + \bm{B} i_{k-1} + \bm{w}_{k-1},
\label{eq:state_model}
\end{equation}

Here, $\bm{A}\in\mathbb{R}^{5\times5}$ is the state-transition matrix, $\bm{B}\in\mathbb{R}^{5\times1}$ is the current-input vector, and $i_{k-1}\triangleq I(t_{k-1})$ is the applied current. The vector $\bm{w}_{k-1}\in\mathbb{R}^{5}$ represents lumped uncertainty and satisfies $\|\bm{w}_{k-1}\|\le\bar w$.

The matrices are specified in Eqs.~\eqref{eq:A_def} and \eqref{eq:B_def}:

\begin{equation}
\label{eq:A_def}
\bm{A} =
\begin{bmatrix}
1 & 0 & 0 & 0 & 0 \\
0 & 1-\dfrac{\Delta t}{\tau_{s,p}} & 0 & 0 & 0 \\
0 & 0 & 1-\dfrac{\Delta t}{\tau_{s,n}} & 0 & 0 \\
0 & 0 & 0 & 1-\dfrac{\Delta t}{\tau_{e}} & 0 \\
0 & 0 & 0 & 0 & 1-\dfrac{\Delta t}{\tau_{e}}
\end{bmatrix},
\end{equation}

\begin{equation}
\label{eq:B_def}
\bm{B} =
\begin{bmatrix}
-\dfrac{\Delta t}{Q_{\mathrm{now}}} \\[2mm]
\dfrac{12D_{p} r^{s}_{p}\Delta t}{7\tau_{s,p}Q_{\mathrm{now}}} \\[2mm]
\dfrac{12D_{n} r^{s}_{n}\Delta t}{7\tau_{s,n}Q_{\mathrm{now}}} \\[2mm]
\dfrac{\Delta t P_{\mathrm{con},a}}{\tau_{e}} \\[2mm]
\dfrac{\Delta t P_{\mathrm{con},b}}{\tau_{e}}
\end{bmatrix}.
\end{equation}

The first row of $\bm{B}$ corresponds to the Coulomb-counting SOC update Eq.~\eqref{eq:soc_cc}. The second and third rows are obtained from the solid-phase concentration-deviation dynamics in Eqs.~\eqref{eq:dxsp_disc}--\eqref{eq:dxsn_disc}, and the last two rows are obtained from the electrolyte concentration dynamics in Eqs.~\eqref{eq:dc1_disc}--\eqref{eq:dc2_disc}.

Unmodeled dynamics and uncertainty are lumped into $\bm{w}_k$. In this paper, $\bm{w}_k$ represents the effect of parameter disturbances, current fluctuation, and unmodeled nonlinearities, and is assumed bounded by $\bar w$.

The measured terminal voltage is written compactly as

\begin{equation}
y_k \triangleq U_{\mathrm{meas}}(t_k)
= h(\bm{x}_k,i_k,\bm{\theta}_k)+v_k,
\label{eq:output_model_compact}
\end{equation}

where \(v_k\) denotes bounded measurement noise. The nonlinear voltage map \(h(\cdot)\) is explicitly defined as

\begin{equation}
h(\bm{x}_k,i_k,\bm{\theta}_k)
= E_{\mathrm{OCV}}(\bm{x}_k,\bm{\theta}_k)
-\eta_{\mathrm{con}}(\bm{x}_k,i_k,\bm{\theta}_k)
-\eta_{\mathrm{act}}(\bm{x}_k,i_k,\bm{\theta}_k)
-\eta_{\mathrm{ohm}}(i_k,\bm{\theta}_k),
\label{eq:nonlinear_voltage_map}
\end{equation}

where the four voltage components correspond to the electrochemical voltage expression in Eq.~\eqref{eq:terminal_voltage}. Accordingly, the estimated output is given by

\begin{equation}
\hat{y}_k \triangleq h(\hat{\bm{x}}_k,i_k,\hat{\bm{\theta}}_k).
\label{eq:estimated_output}
\end{equation}

The measurable output residual is then defined as
\begin{equation}
e_{y,k} \triangleq y_k-\hat{y}_k.
\label{eq:residual_def}
\end{equation}

Before constructing the observer, we clarify the local observability of the reduced-order state vector. Since the terminal voltage is the only measured output, observability is determined by the pair formed by the state dynamics and the nonlinear voltage map, rather than by the state-transition matrix $\bm{A}$ alone. For the locally linearized output map, define

\begin{equation}
\bm{C}_k \triangleq
\left.
\frac{\partial h(\bm{x},i_k,\bm{\theta}_k)}{\partial \bm{x}}
\right|_{\bm{x}=\hat{\bm{x}}_k}.
\label{eq:Ck_def}
\end{equation}

The finite-horizon local observability matrix is then

\begin{equation}
\bm{\mathcal{O}}_{k,N}
=
\begin{bmatrix}
\bm{C}_k\\
\bm{C}_{k+1}\bm{A}\\
\bm{C}_{k+2}\bm{A}^2\\
\vdots\\
\bm{C}_{k+N-1}\bm{A}^{N-1}
\end{bmatrix}.
\label{eq:local_observability_matrix}
\end{equation}

The reduced-order state vector is locally observable along the operating trajectory if $\mathrm{rank}(\bm{\mathcal{O}}_{k,N})=n$. A detailed discussion is provided in Appendix~\ref{app:observability_identifiability}. In particular, for the NMC/graphite cell considered in this work, the OCV curve used in the IESP model does not exhibit a long flat plateau over the operating SOC range, and therefore the terminal voltage retains nonzero local sensitivity to the electrode stoichiometric states. Moreover, the positive and negative solid-phase deviation states are governed by different effective time constants, which prevents structural rank loss of the solid-state observability submatrix under normal operating conditions. The proposed observer is therefore intended for SOC-oriented estimation of the reduced-order IESP states, rather than full reconstruction of distributed electrode concentration profiles.

To ensure online implementability, the switching injection is constructed using the measurable residual $e_{y,k}$ (rather than the unmeasurable state error $\bm{x}_k-\hat{\bm{x}}_k$). The state SMO is formulated as

\begin{equation}
\hat{\bm{x}}_{k}
= \bm{A} \hat{\bm{x}}_{k-1} + \bm{B} i_{k-1}
+ \bm{K} \,\mathrm{sgn}\!\big(e_{y,k-1}\big)
+ \bm{L}\,e_{y,k-1},
\label{eq:state_smo_complete}
\end{equation}

where $\bm{K}\in\mathbb{R}^{n\times 1}$ is the sliding-mode injection gain, $\bm{L}\in\mathbb{R}^{n\times 1}$ is the output-injection gain, and $\mathrm{sgn}(\cdot)$ is the sign function.
(In discrete-time implementation with measurement noise, the discontinuous switching term may induce chattering; in this work, the adaptive dead-zone mechanism in Section~\ref{subsec:adz_lyap} primarily prevents residual fluctuations from propagating into the parameter adaptation loop.)

Let the estimation error be $\tilde{\bm{x}}_k \triangleq \bm{x}_k-\hat{\bm{x}}_k$. 
Combining the plant model Eq.~\eqref{eq:state_model} with the residual-driven observer update Eq.~\eqref{eq:state_smo_complete} gives the following residual-driven error recursion:

\begin{equation}
\tilde{\bm{x}}_k
= \bm{A}\tilde{\bm{x}}_{k-1} + \bm{w}_{k-1}
- \bm{K}\,\mathrm{sgn}\!\big(e_{y,k-1}\big)
- \bm{L}\,e_{y,k-1}.
\label{eq:state_error_dyn}
\end{equation}

Eq.~\eqref{eq:state_error_dyn} is a residual-driven error propagation form used to interpret how the measurable output residual injects corrections; the observer is implemented directly in Eq.~\eqref{eq:state_smo_complete}. Eq.~\eqref{eq:state_error_dyn} highlights how the measurable terminal-voltage residual $e_{y,k}$ enters the estimation-error propagation through the injection terms. 
Importantly, the switching injection is constructed from $e_{y,k}$ for online implementability; we do not require access to the unmeasurable state error $\tilde{\bm{x}}_k$.
In practice, the discontinuous term $\bm{K}\,\mathrm{sgn}(e_{y,k-1})$ provides a high-gain corrective action according to the residual sign, while the linear term $\bm{L} e_{y,k-1}$ improves convergence and tracking. 
The gains $\bm{K}$ and $\bm{L}$ are selected empirically under bounded-disturbance/noise assumptions to balance robustness and transient performance (see Section~4.1 for the values used in this work).

\subsection{Parameter SMO Design}
\label{subsec:param_smo}

The online-adapted parameter subset is selected based on terminal-voltage sensitivity. To support this selection, a sensitivity analysis was conducted under the UDDS condition, as detailed in Appendix~\ref{app:param_sensitivity}. The results show that \(D_p\), \(D_n\), \(Q_{\mathrm{all}}\), \(x_{s,p,0}\), and \(x_{s,n,0}\) have the most pronounced influence on the terminal-voltage response. This finding is also consistent with the sensitivity results reported in our previous electrochemical-modeling study~\cite{guo2025control}. Therefore, online updating is restricted to this compact set of dominant parameters:

\begin{equation}
\bm{\theta}_k \triangleq \begin{bmatrix}
D_{p},\; D_{n},\; Q_{\mathrm{all}},\; x_{s,p,0},\; x_{s,n,0}
\end{bmatrix}^{T}\in\mathbb{R}^{n_\theta}.
\end{equation}

Parameters are assumed to vary slowly:

\begin{equation}
\bm{\theta}_k = \bm{\theta}_{k-1} + \bm{w}^{\theta}_{k-1}, \qquad \|\bm{w}^{\theta}_{k-1}\|\le \bar w_\theta .
\label{eq:param_rw}
\end{equation}

The candidate parameter update law is driven by the measurable voltage residual:

\begin{equation}
\hat{\bm{\theta}}_{k+1}^{\mathrm{cand}}
=
\hat{\bm{\theta}}_{k}
+
\bm{K}^{\theta}\operatorname{sgn}(e_{y,k})
+
\bm{L}^{\theta}e_{y,k},
\label{eq:param_smo}
\end{equation}

where $\bm{K}^{\theta}\in\mathbb{R}^{n_\theta\times 1}$ and $\bm{L}^{\theta}\in\mathbb{R}^{n_\theta\times 1}$ are the parameter-observer gains. Equation~\eqref{eq:param_smo} defines the candidate parameter-SMO update. The actual parameter estimate is then updated conditionally by the adaptive dead-zone gate derived in Section~\ref{subsec:adz_lyap}; when the gate is closed, the parameter estimate is frozen.

\subsection{Lyapunov-Guided Adaptive Dead-Zone for Joint State--Parameter Estimation}
\label{subsec:adz_lyap}

Transient state-estimation errors, measurement noise, and model mismatch can propagate through the voltage residual into the parameter-update branch. The adaptive dead-zone gates this branch while the state observer continues to update. The boundary gain is motivated by a nominal Lyapunov bound; the forcing-aware decrease condition and the implemented gate are distinguished in Appendix~\ref{app:lyapunov_deadzone_proof}.

The state-observer update is written in one-step-ahead form as

\begin{equation}
\hat{\bm{x}}_{k+1}
=\bm{A}\hat{\bm{x}}_k+\bm{B}i_k
+\bm{K}\operatorname{sgn}(e_{y,k})+\bm{L}e_{y,k}.
\label{eq:state_smo_shifted}
\end{equation}

Define the observable state--input scale

\begin{equation}
\begin{aligned}
\hat{\bm{z}}_k
&=\hat{\bm{x}}_k+\bm{A}^{-1}\bm{B}i_k
+\bm{A}^{-1}\bm{K}\phi_k,\\
\phi_k&=\operatorname{sgn}(e_{y,k}).
\end{aligned}
\label{eq:zk_def}
\end{equation}

Here, $|\phi_k|\le1$, and $\bm{A}$ is assumed nonsingular. The nominal output Jacobian $\bm{C}$ defines

\begin{equation}
\bm{A}_c=\bm{A}-\bm{L}\bm{C}.
\label{eq:Ac_const}
\end{equation}

The Lyapunov construction requires $\bm{A}_c$ to be Schur stable and a symmetric positive-definite solution of

\begin{equation}
\bm{A}_c^{\top}\bm{P}\bm{A}_c-\bm{P}=-\bm{Q},
\qquad \bm{Q}\succ0.
\label{eq:lyap_eq}
\end{equation}

Using the positive coefficient ratio defined in Appendix~\ref{app:lyapunov_deadzone_proof}, the nominal online boundary is

\begin{equation}
\delta_k
=\sqrt{\frac{c_z}{c_e}}
\left\|\hat{\bm{x}}_k+\bm{A}^{-1}\bm{B}i_k
+\bm{A}^{-1}\bm{K}\operatorname{sgn}(e_{y,k})\right\|.
\label{eq:law}
\end{equation}

The weighted Young parameter $\beta$ must make $(1+\beta)\bm{Q}-\beta\bm{P}$ positive definite. The candidate value $\beta=10^{-4}$ is admissible only when this condition is satisfied by the actual Lyapunov solution. The state-estimation error bound is established separately in Appendix~\ref{app:state_smo_stability}.

\begin{equation}
g_k=
\begin{cases}
1,& |e_{y,k}|\le\delta_k,\\
0,& |e_{y,k}|>\delta_k.
\end{cases}
\label{eq:adz_gate_signal}
\end{equation}

The gate accepts or freezes the candidate parameter update according to

\begin{equation}
\hat{\bm{\theta}}_{k+1}=
\begin{cases}
\hat{\bm{\theta}}_{k+1}^{\mathrm{cand}},&g_k=1,\\
\hat{\bm{\theta}}_k,&g_k=0.
\end{cases}
\label{eq:adz_gated_param_update}
\end{equation}

Algorithm~\ref{alg:adz} summarizes the conditional design and online update procedure.

\begin{algorithm}[H]
\caption{Adaptive dead-zone gating for joint state--parameter estimation}
\label{alg:adz}
\begin{algorithmic}[1]
\State \textbf{Offline:} Select a nominal Jacobian $\bm{C}$ and gains giving a Schur-stable $\bm{A}_c$. Choose $\bm{Q}\succ0$ and solve Eq.~\eqref{eq:lyap_eq} for $\bm{P}\succ0$.
\State Choose $\beta>0$ such that $(1+\beta)\bm{Q}-\beta\bm{P}\succ0$; use $\beta=10^{-4}$ only if this test passes. Compute $c_z$ and $c_e$ from Eq.~\eqref{eq:app_cz_ce}.
\For{$k=0,1,2,\ldots$}
\State Compute $\hat{y}_k=h(\hat{\bm{x}}_k,i_k,\hat{\bm{\theta}}_k)$ and $e_{y,k}=y_k-\hat{y}_k$.
\State Compute $\delta_k$ using Eq.~\eqref{eq:law}.
\State Update $\hat{\bm{x}}_{k+1}$ using Eq.~\eqref{eq:state_smo_shifted}.
\If{$|e_{y,k}|\le\delta_k$}
\State Accept $\hat{\bm{\theta}}_{k+1}=\hat{\bm{\theta}}_{k+1}^{\mathrm{cand}}$.
\Else
\State Freeze $\hat{\bm{\theta}}_{k+1}=\hat{\bm{\theta}}_k$.
\EndIf
\EndFor
\end{algorithmic}
\end{algorithm}

\section{Results and Discussion}
The reference SOC is calculated by integrating the applied current from the known initial SOC, using the condition-specific effective capacity and the Coulomb-counting relation in Eq.~\eqref{eq:soc_cc}. It is not an observer estimate.

To validate the effectiveness of the proposed ADZ-DSMO, the results are organized into five parts: (i) correct SOC initialization, (ii) incorrect SOC initialization, (iii) computational efficiency, (iv) temperature robustness, and (v) SOC estimation accuracy under model-generated aging conditions. The purpose of the comparisons in this section is not to provide an exhaustive benchmark of all SOC estimators, but to evaluate the effect of the proposed Lyapunov-guided adaptive dead-zone on mitigating error propagation in joint state--parameter estimation. Accordingly, we compare the proposed method with the conventional DSMO, the IESP-based DEKF, and ECM-based benchmarks where applicable. To emulate voltage measurement noise in practical BMS applications, zero-mean Gaussian white noise with a standard deviation of 1 mV was added to the measured terminal-voltage signal in all validation cases. Therefore, all reported SOC estimation results, including the constant-current, dynamic-driving, aging, and temperature-robustness cases, were obtained under noisy voltage measurement conditions.
\subsection{Correct SOC Initialization}

At room temperature ($25^{\circ}\mathrm{C}$), the proposed algorithm is evaluated under four operating conditions, namely 1C constant-current discharge, 3C high-rate constant-current discharge, the UDDS profile, and the NEDC profile. The 3C case is included to examine the observer under high-rate discharge, while the UDDS and NEDC profiles are used to evaluate dynamic driving conditions with variable current excitation. The UDDS and NEDC profiles contain high-rate current transients, with peak discharge rates of approximately 1.83C and 2.06C, respectively. For the Panasonic NCR18650PF cell used in this study, the manufacturer-specified maximum continuous discharge rate is 3C, indicating that all adopted current profiles remain within the allowable operating range.

In this subsection, the SOC is correctly initialized to 1. The initial values of the observer states and parameter estimator are set according to the identified parameters listed in Table~\ref{tab:params}. The sampling time is set to $\Delta t = 1$~s. The proposed ADZ-DSMO is compared with the conventional IESP-based DSMO and the IESP-based DEKF baseline.

For the IESP-based SMO methods, the observer gain vectors used in the simulations were set as follows:

\[
\begin{aligned}
\bm{L} &=
\begin{bmatrix}
2.0\times10^{-3} &
1.0\times10^{-6} &
1.0\times10^{-6} &
2.0\times10^{-1} &
2.0\times10^{-1}
\end{bmatrix}^{T},\\
\bm{K} &=
\begin{bmatrix}
5.0\times10^{-3} &
2.5\times10^{-6} &
2.5\times10^{-6} &
2.5\times10^{-1} &
2.5\times10^{-1}
\end{bmatrix}^{T},\\
\bm{L}^{\theta} &=
\begin{bmatrix}
5.0\times10^{-4} &
-5.0\times10^{-4} &
5.0\times10^{-1} &
5.0\times10^{-4} &
-5.0\times10^{-4}
\end{bmatrix}^{T},\\
\bm{K}^{\theta} &=
\begin{bmatrix}
2.5\times10^{-6} &
2.5\times10^{-6} &
2.5\times10^{-3} &
2.5\times10^{-6} &
2.5\times10^{-6}
\end{bmatrix}^{T}.
\end{aligned}
\]

For the IESP-based DEKF, the process-noise covariance matrices of the state estimator and the parameter estimator were set as

\begin{displaymath}
\begin{aligned}
\bm{Q}^{x}
&=
\operatorname{diag}\!\left(
5.0\times 10^{-3},
2.5\times 10^{-5},
2.5\times 10^{-5},
2.5\times 10^{-1},
2.5\times 10^{-1}
\right),\\
\bm{Q}^{\theta}
&=
\operatorname{diag}\!\left(
5.0\times 10^{-6},
5.0\times 10^{-6},
2.5\times 10^{-2},
5.0\times 10^{-6},
5.0\times 10^{-6}
\right).
\end{aligned}
\end{displaymath}

The measurement-noise covariance values for the state and parameter estimators were both set to

\begin{displaymath}
\bm{R}^{x}=\bm{R}^{\theta}=1.
\end{displaymath}

Observer settings are unchanged across the validation scenarios.

\begin{figure}[H] \centering \includegraphics[width=\linewidth]{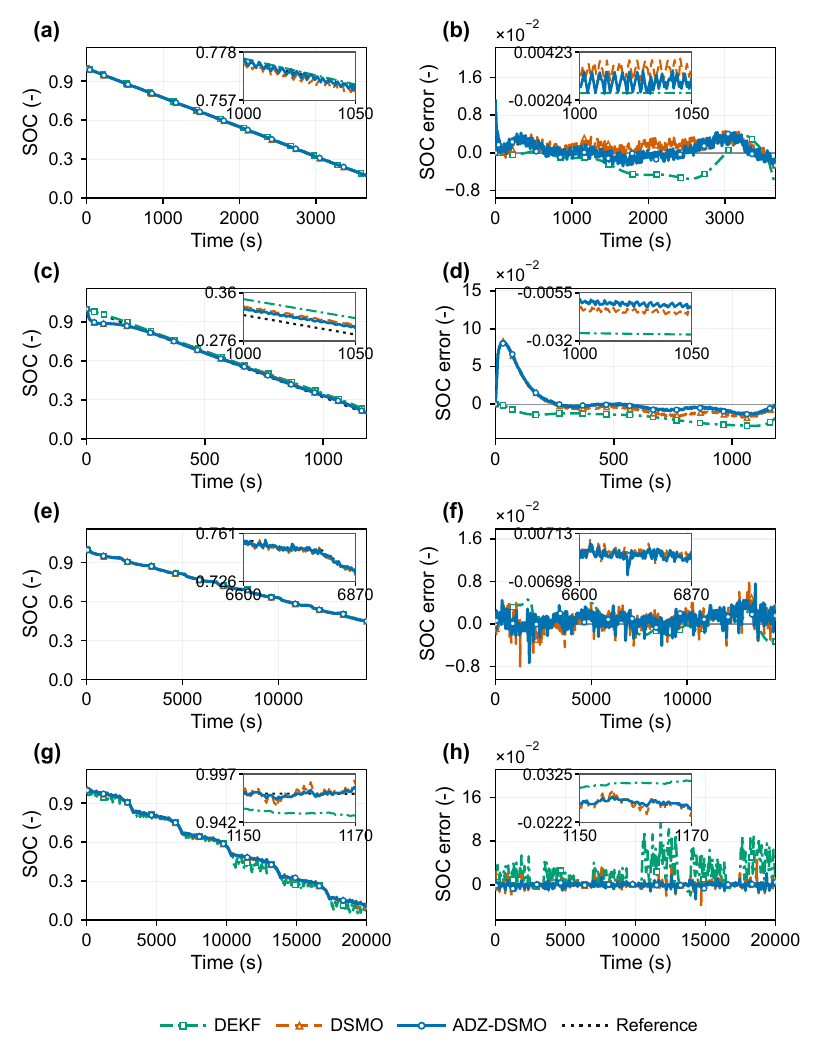} \caption{SOC estimation with correct initialization ($\mathrm{SOC}_0=1$): (a,b) 1C; (c,d) 3C; (e,f) UDDS; (g,h) NEDC. Left panels: SOC; right panels: SOC error. Both quantities are dimensionless.} \label{fig:soc_estimation_no_init_error} \end{figure}

\begin{table}[H]
\centering
\footnotesize
\caption{SOC estimation errors of each algorithm with correct SOC initialization}
\label{tab:SOC_errors}
\begin{adjustbox}{max width=\linewidth}
\begin{tabular}{l l c}
\toprule
\textbf{Working condition} & \textbf{Algorithm} & \textbf{MAE (\%)} \\
\midrule
UDDS & ADZ-DSMO  & 0.14 \\
UDDS & DSMO                     & 0.16 \\
UDDS & DEKF                     & 0.16 \\
UDDS & EKF with ECM~\cite{guo2018parameter} & 1.08 \\
UDDS & FDZ-DEKF with ECM~\cite{guo2018parameter} & 0.23 \\
\midrule
1C   & ADZ-DSMO  & 0.13 \\
1C   & DSMO                     & 0.18 \\
1C   & DEKF                     & 0.25 \\
1C   & EKF with ECM~\cite{guo2018parameter} & 2.79 \\
1C   & FDZ-DEKF with ECM~\cite{guo2018parameter} & 0.39 \\
\midrule
3C   & ADZ-DSMO  & 1.13 \\
3C   & DSMO                     & 1.47 \\
3C   & DEKF                     & 1.75 \\
\midrule
NEDC & ADZ-DSMO  & 0.21 \\
NEDC & DSMO                     & 0.34 \\
NEDC & DEKF                     & 2.05 \\
\bottomrule
\end{tabular}
\end{adjustbox}
\end{table}

Figure~\ref{fig:soc_estimation_no_init_error} compares the SOC trajectories and estimation errors under correct initialization. The proposed ADZ-DSMO closely follows the reference SOC under the tested operating conditions. In the error plots, the proposed method maintains a narrower error band and suppresses error fluctuations more effectively than the conventional DSMO and the DEKF. 

The quantitative results in Table~\ref{tab:SOC_errors} are consistent with the graphical observations. The proposed ADZ-DSMO achieves the lowest MAE under all tested conditions, with errors of 0.14\%, 0.13\%, 1.13\%, and 0.21\% under UDDS, 1C, 3C, and NEDC, respectively. Compared with the conventional DSMO, the MAE is reduced from 0.16\% to 0.14\% under UDDS, from 0.18\% to 0.13\% under 1C, from 1.47\% to 1.13\% under 3C, and from 0.34\% to 0.21\% under NEDC. Compared with the DEKF, the proposed method also achieves lower MAE under all four profiles. Although the SOC error increases under the 3C high-rate condition because of stronger polarization and model mismatch, the proposed method still provides the best accuracy among the compared IESP-based observers. In addition, compared with the ECM-based methods reported in~\cite{guo2018parameter} where applicable, the electrochemical-model-based algorithms achieve substantially lower errors, confirming the benefit of incorporating simplified electrochemical dynamics into the SOC observer.

\subsection{Incorrect SOC Initialization} 
\begin{figure}[H]
    \centering
    \includegraphics[width=\linewidth]{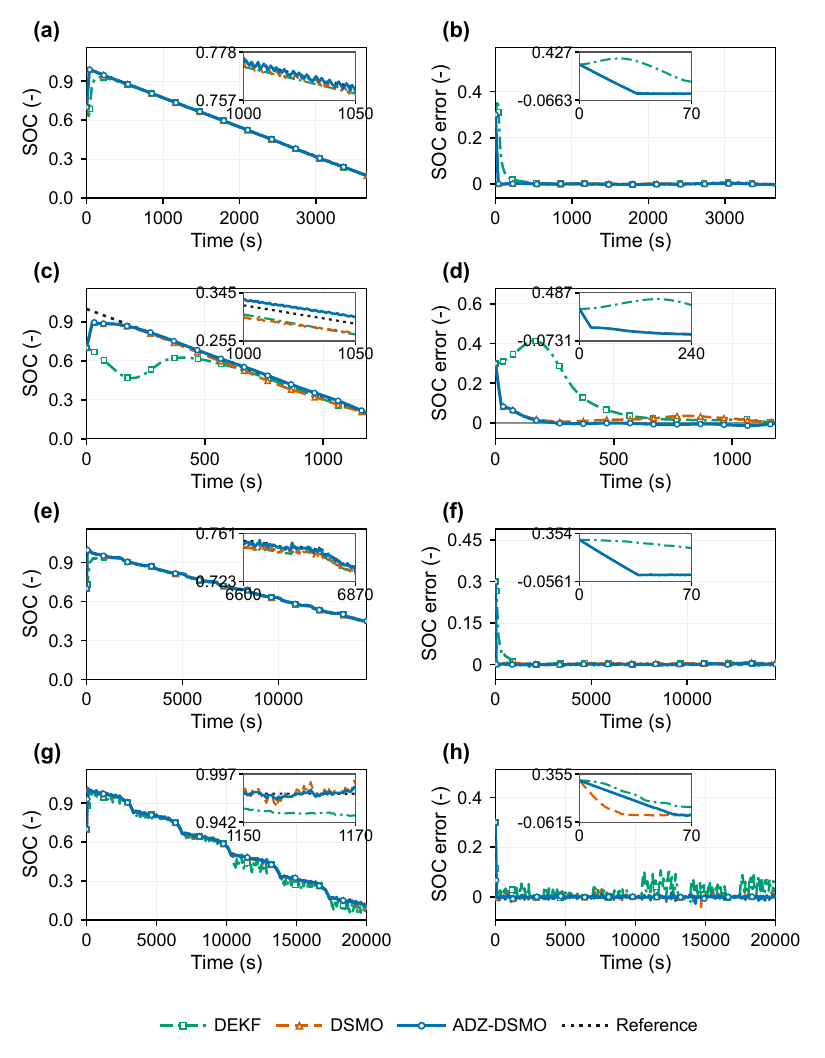}
\caption{SOC estimation with a 30\% initial error ($\mathrm{SOC}_0=0.7$): (a,b) 1C; (c,d) 3C; (e,f) UDDS; (g,h) NEDC. Left panels: SOC; right panels: SOC error. Error insets resolve early recovery. Line and marker conventions match Figure~\ref{fig:soc_estimation_no_init_error}.} 
    \label{fig:soc_estimation_init_error}
\end{figure}

The same four operating conditions are evaluated with an initial SOC estimate of 0.7, corresponding to a 30\% initial error. Observer gains and initial parameter values are unchanged from the correct-initialization case. The convergence time is defined as the first sampling instant at which the absolute SOC estimation error enters the $\pm1\%$ error band. With $\Delta t=1$~s, convergence time is reported in seconds. The corresponding SOC trajectories, estimation errors, MAE values, and convergence times are shown in Figure~\ref{fig:soc_estimation_init_error} and Table~\ref{tab:SOC_errors_in}.

\begin{table}[H]
\centering
\footnotesize
\caption{SOC estimation errors and convergence time of each algorithm with 30\% initial SOC error}
\label{tab:SOC_errors_in}
\renewcommand{\arraystretch}{1.08}
\setlength{\tabcolsep}{4pt}
\footnotesize
\begin{tabularx}{\linewidth}{l X c c}
\toprule
\makecell{\textbf{Working}\\\textbf{condition}} 
& \textbf{Algorithm} 
& \textbf{MAE (\%)} 
& \makecell{\textbf{Convergence}\\\textbf{time (s)}} \\
\midrule
UDDS & ADZ-DSMO  & 0.17 & 36 \\
UDDS & DSMO                     & 0.39 & 36 \\
UDDS & DEKF                     & 0.69 & 920 \\
UDDS & EKF with ECM~\cite{guo2018parameter} & 1.28 & 100 \\
UDDS & FDZ-DEKF with ECM~\cite{guo2018parameter} & 0.74 & 100 \\
\midrule
1C   & ADZ-DSMO  & 0.13 & 35 \\
1C   & DSMO                     & 0.18 & 35 \\
1C   & DEKF                     & 0.95 & 379 \\
1C   & EKF with ECM~\cite{guo2018parameter} & 3.24 & 100 \\
1C   & FDZ-DEKF with ECM~\cite{guo2018parameter} & 2.05 & 100 \\
\midrule
3C   & ADZ-DSMO  & 1.46 & 187 \\
3C   & DSMO                     & 2.67 & 217 \\
3C   & DEKF                     & 11.98 & 1116 \\
\midrule
NEDC & ADZ-DSMO  & 0.25 & 26 \\
NEDC & DSMO                     & 0.36 & 26 \\
NEDC & DEKF                     & 2.12 & 185 \\
\bottomrule
\end{tabularx}
\end{table}

Figure~\ref{fig:soc_estimation_init_error} shows rapid initial-error recovery by the proposed method and a narrower post-recovery error band than that of DSMO. DEKF exhibits slower error decay and larger residual errors, particularly under UDDS, 3C, and NEDC.

Table~\ref{tab:SOC_errors_in} reports MAEs of 0.17\%, 0.13\%, 1.46\%, and 0.25\% for the proposed method under UDDS, 1C, 3C, and NEDC, respectively. The corresponding values are 0.39\%, 0.18\%, 2.67\%, and 0.36\% for DSMO, and 0.69\%, 0.95\%, 11.98\%, and 2.12\% for DEKF. The 3C comparison highlights the benefit of selective parameter adaptation when a large initial mismatch coincides with stronger polarization.

The proposed method reaches the $\pm1\%$ error band in 36~s under UDDS, 35~s under 1C, 187~s under 3C, and 26~s under NEDC; DEKF requires 920, 379, 1116, and 185~s, respectively. The similar recovery times under UDDS and 1C reflect the dominant initial mismatch and residual-driven SMO correction. Stronger polarization and model mismatch delay recovery under 3C, but the proposed method still converges faster than DSMO (217~s) and DEKF.

DSMO provides a within-family baseline for isolating the gate effect, while the IESP-based DEKF provides a dual-filtering benchmark. A state-only UKF improves nonlinear state propagation but does not regulate the coupled parameter-update loop. Adaptive or super-twisting state SMOs primarily address robust correction and chattering. The present comparisons focus on SOC accuracy and initial-error recovery in joint state--parameter estimation.

\subsection{Computational Efficiency}

The computational efficiency of the compared observers is evaluated using the average computation time per sampling step. For each algorithm and each operating condition, the complete operating profile is executed 10 times under the same software and hardware settings. In each run, the total wall-clock execution time of the full profile is recorded and then divided by the number of sampling steps to obtain the average per-step computation time. The reported value is the mean and standard deviation of these 10 average per-step computation times. Therefore, the reported computation time represents the average time required to complete one full estimator update at a single sampling instant, including voltage evaluation, state update, dead-zone boundary evaluation, and parameter-update calculations when the parameter estimator is activated.

All algorithms are implemented in MATLAB R2020a and executed on a computer equipped with an Intel\textsuperscript{\textregistered} Core\textsuperscript{TM} i5-9300H CPU and 16~GB DDR4 memory. The sampling period is $\Delta t=1$~s. Two operating conditions are considered at $25^{\circ}\mathrm{C}$, namely 1C constant-current discharge and the UDDS dynamic profile. For each condition, two initialization cases are evaluated: correct SOC initialization and 30\% initial SOC error. In addition to computation time, the running ratio of the parameter estimator is reported to quantify the gating behavior introduced by the adaptive dead-zone mechanism.

For the proposed method, the parameter-estimator running ratio is calculated from the gate signal as
\begin{equation}
\eta_{\mathrm{run}} =
\frac{1}{N}\sum_{k=1}^{N} g_k \times 100\%,
\end{equation}
where $N$ is the number of sampled data points and $g_k$ is the gate signal at time step $k$. Here, $g_k=1$ indicates that the parameter estimator is activated, whereas $g_k=0$ indicates that the parameter update is suppressed by the adaptive dead-zone. Since the sampling period is fixed at 1~s, this sample-based ratio is equivalent to the time-based running ratio.

\begin{figure}[H]
    \centering
    \includegraphics[width=\linewidth]{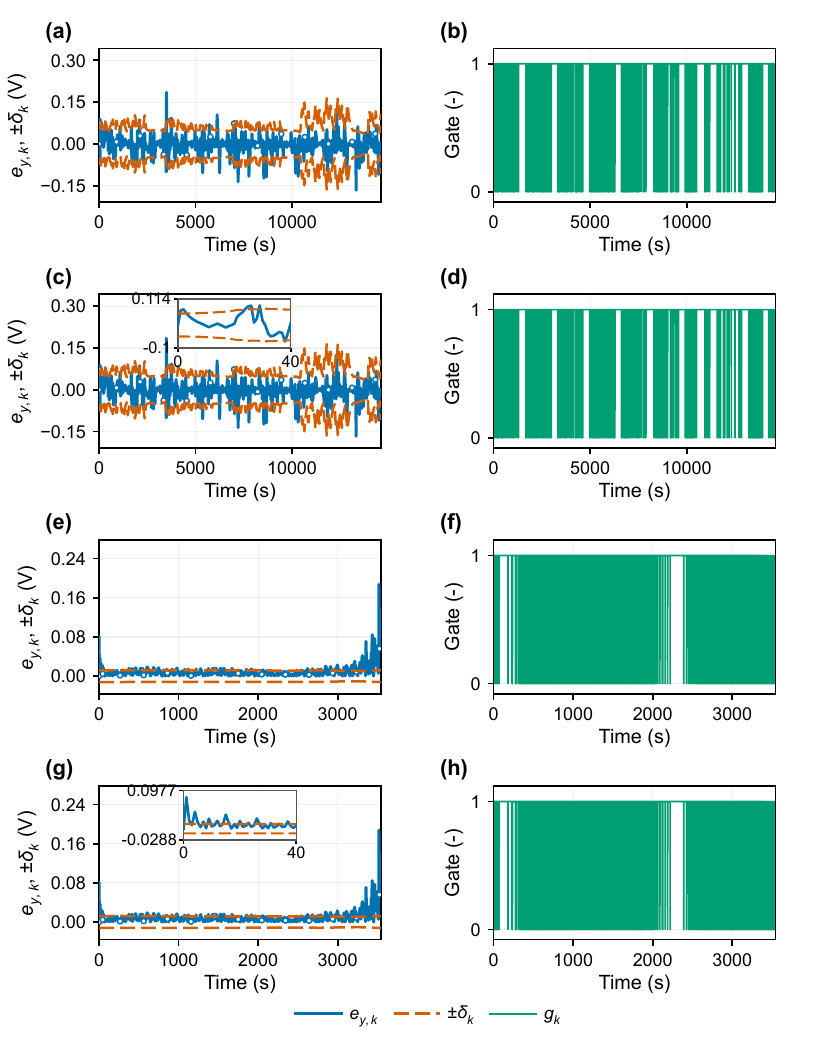}
    \caption{Adaptive boundary, voltage residual, and gate: (a,b) UDDS, correct initialization; (c,d) UDDS, 30\% initial error; (e,f) 1C, correct initialization; (g,h) 1C, 30\% initial error. Orange boundaries: $\pm\delta_k$; blue curve: $e_{y,k}$. Gate 1 enables parameter estimation; gate 0 suspends it.}
    \label{fig:deadzone_gate_trends}
\end{figure}

\begin{table}[H]
\centering
\footnotesize
\caption{Computation time and parameter-estimator running ratio of different SOC estimation algorithms}
\label{tab:comp_time}
\renewcommand{\arraystretch}{1.08}
\setlength{\tabcolsep}{3pt}
\footnotesize
\begin{adjustbox}{max width=\linewidth}
\begin{tabular}{l l c c c c}
\toprule
\makecell{\textbf{Working}\\\textbf{condition}} &
\makecell{\textbf{Initial}\\\textbf{SOC}} &
\makecell{\textbf{DEKF}\\\textbf{(ms)}} &
\makecell{\textbf{DSMO}\\\textbf{(ms)}} &
\makecell{\textbf{ADZ-DSMO}\\\textbf{(ms)}} &
\makecell{\textbf{Parameter-estimator}\\\textbf{running ratio}\\\textbf{(\%)}} \\
\midrule
UDDS & No initial error
& $18.03 \pm 0.54$
& $15.15 \pm 0.43$
& $13.26 \pm 0.39$
& 90.11 \\

UDDS & 30\% initial error
& $17.84 \pm 0.52$
& $14.26 \pm 0.41$
& $13.42 \pm 0.40$
& 90.11 \\

1C & No initial error
& $17.85 \pm 0.51$
& $14.23 \pm 0.40$
& $11.25 \pm 0.34$
& 75.20 \\

1C & 30\% initial error
& $17.57 \pm 0.50$
& $14.04 \pm 0.39$
& $10.95 \pm 0.33$
& 75.20 \\
\bottomrule
\end{tabular}
\end{adjustbox}

\vspace{1mm}
\footnotesize{The computation time is reported as mean $\pm$ standard deviation of the average per-step computation time obtained from 10 full-profile runs.}
\end{table}

Figure~\ref{fig:deadzone_gate_trends} illustrates the dynamic dead-zone boundary, the voltage residual, and the corresponding gate signal of the proposed method. The dead-zone boundary is updated online rather than fixed. When the residual remains within the selected adaptive boundary, i.e., $|e_{y,k}|\leq\delta_k$, the gate signal is set to $g_k=1$ and the parameter estimator is activated. When the residual exceeds this boundary, i.e., $|e_{y,k}|>\delta_k$, the gate signal becomes $g_k=0$ and the parameter update is skipped. Therefore, the proposed method selectively activates the parameter estimator only when the residual satisfies the stability-guided update condition.

The gate statistics in Table~\ref{tab:comp_time} directly quantify this selective-update behavior. Under UDDS, the parameter estimator is active for 90.11\% of the samples, whereas under 1C constant-current discharge the running ratio decreases to 75.20\%. This difference is consistent with the excitation characteristics of the two profiles. The UDDS profile contains stronger dynamic current excitation and more frequent residual variations, which lead to more frequent parameter-estimator activation. In contrast, the 1C constant-current profile provides weaker excitation for parameter identification; therefore, the adaptive dead-zone suppresses a larger fraction of unnecessary parameter updates.

The computation-time results in Table~\ref{tab:comp_time} show that the proposed ADZ-DSMO is computationally more efficient than both the DEKF and the conventional DSMO. Under UDDS with correct initialization, the average computation time decreases from 18.03~ms for the DEKF to 13.26~ms for the proposed method, corresponding to a reduction of approximately 26.5\%. Under 1C with correct initialization, the time decreases from 17.85~ms to 11.25~ms, corresponding to a reduction of approximately 37.0\%. Similar reductions are observed in the 30\% initial-error cases.

Compared with the conventional DSMO, the proposed method also reduces the computation time by avoiding unnecessary parameter-estimator updates. Under 1C with a 30\% initial SOC error, the average per-step computation time decreases from 14.04~ms for the DSMO to 10.95~ms for the proposed method, corresponding to a reduction of approximately 22.0\%. Under 1C with correct initialization, the reduction is approximately 20.9\%. The reduction is smaller under UDDS because the parameter estimator remains active for a larger fraction of the profile, as indicated by the 90.11\% running ratio. Nevertheless, even under the UDDS dynamic profile, the proposed method remains faster than both the DEKF and the conventional DSMO.

Overall, selective parameter updating balances SOC accuracy and computational efficiency. The absolute times depend on the software and hardware platform, while the identical MATLAB and hardware settings provide a consistent basis for comparing relative computational demand.

\subsection{Temperature Robustness Analysis}

To further evaluate the robustness of the proposed observer under temperature variations, comparative tests are conducted under the 1C constant-current condition at three ambient temperatures: 15~$^\circ$C, 25~$^\circ$C, and 45~$^\circ$C. The conventional DSMO and the proposed ADZ-DSMO are evaluated using the same reference SOC calculated by current integration. The 1C condition is selected in this analysis to isolate the influence of temperature from the excitation complexity of dynamic driving profiles. 

The three temperatures represent reduced-, room-, and elevated-temperature operation within a practical BMS-controlled range. Sub-zero tests are outside the available dataset and remain a direction for future work.

\begin{figure}[H]
    \centering
    \includegraphics[width=\linewidth]{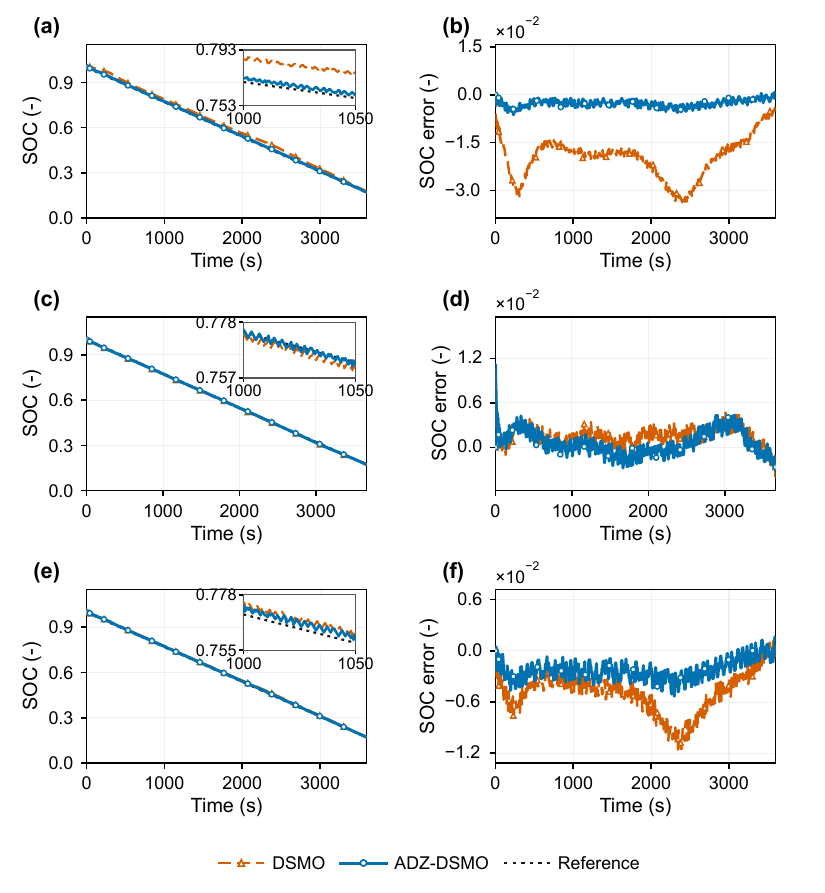}
    \caption{1C SOC estimation at (a,b) $15\,^{\circ}\mathrm{C}$, (c,d) $25\,^{\circ}\mathrm{C}$, and (e,f) $45\,^{\circ}\mathrm{C}$. Left panels: SOC; right panels: SOC error. Both quantities are dimensionless.}
    \label{fig:temperature_results}
\end{figure}

Figure~\ref{fig:temperature_results} shows the SOC estimation trajectories and the corresponding SOC errors at different temperatures. Both observers can track the reference SOC over the full discharge process. However, the proposed ADZ-DSMO exhibits smaller error fluctuations under off-nominal temperature conditions. This advantage is particularly evident at 15~$^\circ$C, where stronger polarization and increased model uncertainty make SOC estimation more challenging. At 25~$^\circ$C, both methods achieve relatively small errors, while the proposed method still provides a modest improvement. At 45~$^\circ$C, the adaptive dead-zone mechanism also reduces the estimation error compared with the conventional DSMO.

The quantitative comparison is summarized in Table~\ref{tab:temperature_results}. The proposed ADZ-DSMO consistently achieves lower MAE than the conventional DSMO at all tested temperatures. Specifically, the MAE is reduced from 1.97\% to 0.28\% at 15~$^\circ$C, from 0.18\% to 0.13\% at 25~$^\circ$C, and from 0.49\% to 0.23\% at 45~$^\circ$C. 
These results indicate that the adaptive dead-zone mechanism improves the temperature robustness of the electrochemical-model-based SOC observer over the tested practical operating-temperature range, especially at the reduced-temperature point of 15~$^\circ$C where model uncertainty and polarization effects become more pronounced.

\begin{table}[H]
\centering
\footnotesize
\caption{SOC estimation errors under different temperatures in the 1C condition}
\label{tab:temperature_results}
\renewcommand{\arraystretch}{1.08}
\setlength{\tabcolsep}{5pt}
\footnotesize
\begin{adjustbox}{max width=\linewidth}
\begin{tabular}{c l c}
\toprule
\textbf{Temperature} & \textbf{Algorithm} & \textbf{MAE (\%)} \\
\midrule
15~$^\circ$C & DSMO & 1.97 \\
15~$^\circ$C & ADZ-DSMO & 0.28 \\
\midrule
25~$^\circ$C & DSMO & 0.18 \\
25~$^\circ$C & ADZ-DSMO & 0.13 \\
\midrule
45~$^\circ$C & DSMO & 0.49 \\
45~$^\circ$C & ADZ-DSMO & 0.23 \\
\bottomrule
\end{tabular}
\end{adjustbox}
\end{table}

\subsection{SOC Estimation Accuracy under Model-Generated Aging Conditions}

Battery aging changes the effective capacity and electrochemical characteristics of lithium-ion cells, which can affect SOC estimation by modifying the relationship between current integration, terminal-voltage response, and SOC evolution. If an observer is calibrated using fresh-cell parameters only, the mismatch between the nominal model and the aged-cell behavior may lead to accumulated SOC estimation errors. Since this work focuses on SOC estimation rather than SOH estimation, this subsection evaluates whether the proposed observer can remain stable and accurate when aging-related parameter variation is introduced in a controlled model-based scenario.

To construct the aging scenarios, the cycle-aging model reported in~\cite{wang2011cycle} is used to calculate the capacity degradation after different cumulative cycling levels. The nominal capacity of the fresh cell is 2894.1~mAh. According to this aging model, the SOH values after 100 and 400 cycles are 97.0\% and 90.0\%, corresponding to effective capacities of 2807.3~mAh and 2604.7~mAh, respectively. These capacity values are introduced into the IESP model to generate the aged terminal-voltage responses under UDDS. The reference SOC is calculated by integrating the applied current using the aged effective capacity, rather than the fresh-cell nominal capacity. The generated terminal voltage is used as the measurement input for the observers. Therefore, this subsection should be interpreted as a model-generated aging robustness test under capacity degradation, rather than an experimental validation using physically aged cells.

\begin{figure}[H]
    \centering
    \includegraphics[width=\linewidth]{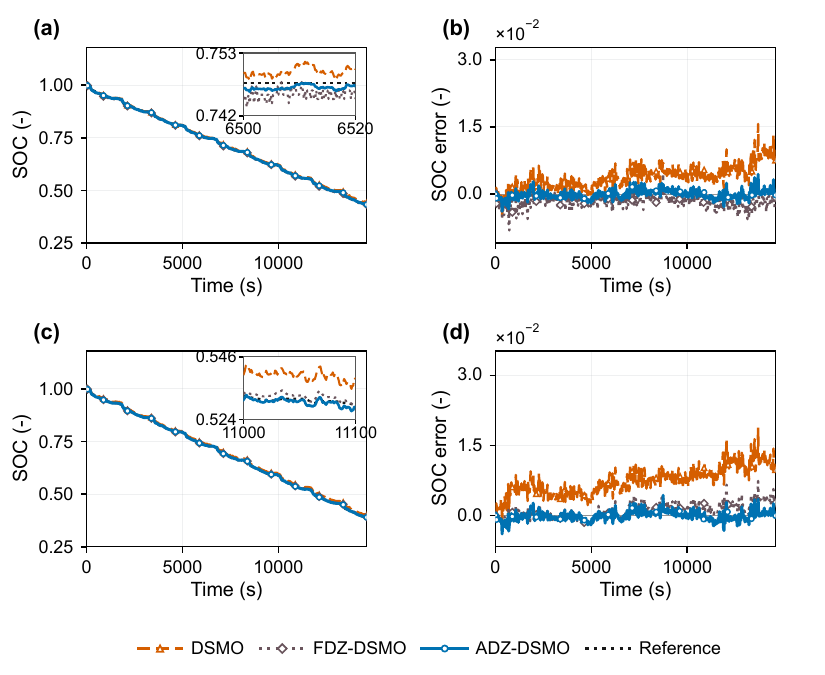}
    \caption{UDDS SOC estimation under capacity-based aging after (a,b) 100 cycles and (c,d) 400 cycles. Left panels: SOC; right panels: SOC error. Both quantities are dimensionless.}
    \label{fig:soc_estimation_aging}
\end{figure}

Three methods are compared under the UDDS profile: the conventional DSMO, the fixed dead-zone dual sliding-mode observer (FDZ-DSMO), and the proposed ADZ-DSMO. The fixed dead-zone interval is set to $(0,0.001)$ following~\cite{guo2018parameter}, whereas the proposed adaptive dead-zone boundary is updated online according to the Lyapunov-guided condition in Eq.~\eqref{eq:law}. The corresponding SOC estimation results are shown in Figure~\ref{fig:soc_estimation_aging}, and the MAE and maximum error values are summarized in Table~\ref{tab:ag_udds_cycle_mae}. The SOH and capacity columns indicate the aging levels used to generate the model-based aged-cell data.

\begin{table}[H]
\centering
\footnotesize
\caption{SOC estimation error under the UDDS condition at different cumulative cycling levels}
\label{tab:ag_udds_cycle_mae}
\renewcommand{\arraystretch}{1.08}
\setlength{\tabcolsep}{3.5pt}
\footnotesize
\begin{tabularx}{\linewidth}{c c c X c c}
\toprule
\textbf{Condition} & \textbf{SOH (\%)} & \makecell{\textbf{Effective}\\\textbf{capacity (mAh)}} 
& \textbf{Algorithm} & \textbf{MAE (\%)} & \makecell{\textbf{Max error}\\\textbf{(\%)}} \\
\midrule
\multirow{3}{*}{100 cycles}
& \multirow{3}{*}{97.0}
& \multirow{3}{*}{2807.3}
& DSMO & 0.38 & 1.57 \\
& & & FDZ-DSMO & 0.18 & 0.81 \\
& & & ADZ-DSMO & 0.08 & 0.44 \\
\midrule
\multirow{3}{*}{400 cycles}
& \multirow{3}{*}{90.0}
& \multirow{3}{*}{2604.7}
& DSMO & 0.73 & 1.88 \\
& & & FDZ-DSMO & 0.14 & 0.74 \\
& & & ADZ-DSMO & 0.08 & 0.43 \\
\bottomrule
\end{tabularx}
\end{table}

As shown in Figure~\ref{fig:soc_estimation_aging}, all three observers can track the current-integrated SOC reference based on the aged capacity under the UDDS profile. This is expected because the aging data are generated in a controlled model-based setting using the same reduced-order electrochemical modeling framework. Therefore, the absolute SOC errors are relatively small. The purpose of this analysis is not to claim full experimental aging validation, but to examine whether the observers remain stable when capacity degradation is introduced into the simulated aged-cell response.

The conventional DSMO shows larger estimation deviations as the degradation level increases. When the SOH decreases from 97.0\% to 90.0\%, the MAE of the DSMO increases from 0.38\% to 0.73\%, and the maximum error increases from 1.57\% to 1.88\%. This indicates that continuous parameter updating without an appropriate gating mechanism can become more sensitive to aging-induced model mismatch.

Both dead-zone-based methods improve estimation robustness under the model-generated aging conditions. At 100 cycles, the FDZ-DSMO reduces the MAE from 0.38\% to 0.18\%, while the proposed ADZ-DSMO further reduces it to 0.08\%. At 400 cycles, the MAE decreases from 0.73\% for the conventional DSMO to 0.14\% for the FDZ-DSMO and 0.08\% for the proposed ADZ-DSMO. Similar trends are observed in the maximum error, where the proposed method keeps the peak error below 0.45\% in both aging cases.

Compared with the FDZ-DSMO, the proposed ADZ-DSMO achieves lower MAE and maximum error at both degradation levels. The improvement is attributed to the online adaptation of the dead-zone boundary. Unlike the fixed dead-zone method, which relies on a manually selected constant threshold, the proposed method updates the admissible residual bound according to the Lyapunov-guided condition. This allows the parameter update to be regulated according to the instantaneous residual behavior, reducing unnecessary or unreliable parameter corrections when aging-related model mismatch is present.

Overall, the proposed method remains accurate in the capacity-degradation scenarios defined by the reference aging model. Validation using physically aged cells and extension toward SOH estimation remain directions for future work.

\section{Conclusions} \label{sec:conclusion} 

This paper developed a Lyapunov-guided ADZ-DSMO for reliable SOC estimation using an improved electrochemical single-particle model. The main contribution is not the dead-zone concept itself, but its stability-guided implementation as an online parameter-update gate within an electrochemical-model-based DSMO. The proposed gate freezes parameter adaptation when the terminal-voltage residual exceeds the Lyapunov-guided boundary and enables joint state--parameter estimation when the residual remains within the selected gate region. Validation under 1C, 3C, UDDS, and NEDC conditions showed that the proposed method achieves lower SOC estimation errors than the conventional DSMO, the IESP-based DEKF, and ECM-based benchmarks where applicable. Under correct SOC initialization, the MAE remained below 0.25\% under 1C, UDDS, and NEDC, and reached 1.13\% under the more demanding 3C condition. Under a 30\% initial SOC error, the proposed observer converged within 26--36~s under 1C, UDDS, and NEDC, and within 187~s under 3C, while maintaining lower post-convergence errors than the compared methods. The computation-time analysis further showed that the adaptive gate reduces unnecessary parameter-estimator updates, leading to lower average per-step computation time than the always-updating DSMO.

The proposed method also showed improved robustness under temperature variation and model-generated aging conditions. In the temperature tests, it consistently reduced SOC estimation errors at 15~$^\circ$C, 25~$^\circ$C, and 45~$^\circ$C compared with the conventional DSMO. In the aging analysis, capacity degradation was introduced using a literature cycle-aging model, and the results should therefore be interpreted as a controlled robustness test under synthetic aging-related parameter variation rather than experimental validation on physically aged cells. Overall, the proposed ADZ-DSMO provides an effective balance between accuracy, robustness, and computational efficiency for electrochemical-model-based SOC estimation. 
A current limitation is that the experimental validation is conducted on a single NMC cell. 
The temperature validation covers a broad practical BMS-controlled operating range from 15~$^\circ$C to 45~$^\circ$C, corresponding to reduced-, room-, and elevated-temperature operation in real-world battery applications with thermal management, but it does not include sub-zero or other extreme low-temperature conditions. In addition, the aging analysis uses model-generated aged voltage responses and a current-integrated SOC reference based on capacities obtained from the cited cycle-aging model; therefore, it should be interpreted as a controlled model-dependent robustness test under synthetic aging-related capacity degradation, rather than experimental validation using physically aged cells. Future work will focus on validation using real aged cells, extension to other battery chemistries after chemistry-specific OCV characterization and parameter calibration, evaluation under more severe low-temperature conditions when suitable experimental data become available, and benchmarking against regularized electrochemical-model-based Dual UKF or other advanced adaptive filtering and observer baselines. In addition, the integration of the proposed Lyapunov-guided adaptive gate with nonlinear filtering frameworks and adaptive or super-twisting DSMO structures will be investigated to further improve the robustness of online state--parameter estimation.

\section*{Acknowledgments}
This work was supported by the Chengdu Technological Innovation Research and Development Project [Grant No.2024-YF08--00031-GX]; the Sichuan Provincial Department of Science and Technology [Grant No.23YYJCYJ0096, Grant No.2024YFHZ0267]; the Fujian Provincial Department of Science and Technology Projects [Grant no: 2025HZ01230051]; and the Research Foundation -- Flanders (FWO) [Grant No. 1252326N].

\section*{Conflict of Interest}
The authors declare that they have no conflict of interest.

\section*{Author Contributions}

\textbf{Guangdi Hu}: Conceptualization, Methodology, Writing -- original draft.  
\textbf{Keyi Liao}: Software, Visualization, Writing -- original draft.  
\textbf{Jian Ye}: Supervision, Writing -- review \& editing.  
\textbf{Feng Guo}: Conceptualization, Supervision, Writing -- original draft, Writing -- review \& editing.

\appendix

\renewcommand{\thesection}{\Alph{section}}
\renewcommand{\theequation}{\thesection.\arabic{equation}}
\renewcommand{\thefigure}{\thesection.\arabic{figure}}
\renewcommand{\thetable}{\thesection.\arabic{table}}

\setcounter{section}{0}
\setcounter{equation}{0}
\setcounter{figure}{0}
\setcounter{table}{0}

\clearpage
\setcounter{equation}{0}
\setcounter{figure}{0}
\setcounter{table}{0}

\refstepcounter{section}
\section*{Appendix \thesection. Local Observability and Practical Identifiability Analysis}
\label{app:observability_identifiability}
\addcontentsline{toc}{section}{Appendix \thesection. Local Observability and Practical Identifiability Analysis}

This appendix discusses the local observability of the reduced-order IESP state vector and the practical identifiability of the online-estimated parameter subset. The purpose is to clarify whether the coupled positive and negative electrode concentration-deviation states compromise the proposed observer design.

For the discrete-time state equation

\begin{equation}
\bm{x}_k = \bm{A} \bm{x}_{k-1}+\bm{B} i_{k-1}+\bm{w}_{k-1},
\end{equation}

and the nonlinear terminal-voltage output

\begin{equation}
y_k=h(\bm{x}_k,i_k,\bm{\theta}_k)+v_k,
\end{equation}

local observability is assessed through the trajectory-dependent linearization

\begin{equation}
\delta \bm{x}_{k+1}=\bm{A}\delta \bm{x}_k,\qquad
\delta y_k=\bm{C}_k\delta \bm{x}_k,
\end{equation}

where

\begin{equation}
\bm{C}_k =
\left.
\frac{\partial h(\bm{x},i_k,\bm{\theta}_k)}{\partial \bm{x}}
\right|_{\bm{x}=\hat{\bm{x}}_k}.
\end{equation}

The finite-horizon local observability matrix is

\begin{equation}
\bm{\mathcal{O}}_{k,N}
=
\begin{bmatrix}
\bm{C}_k\\
\bm{C}_{k+1}\bm{A}\\
\bm{C}_{k+2}\bm{A}^2\\
\vdots\\
\bm{C}_{k+N-1}\bm{A}^{N-1}
\end{bmatrix}.
\label{eq:app_observability_matrix}
\end{equation}

The reduced-order state vector is locally observable over the horizon \(N\) if

\begin{equation}
\mathrm{rank}\left(\bm{\mathcal{O}}_{k,N}\right)=5.
\label{eq:app_observability_rank}
\end{equation}

To examine the electrode concentration-deviation coupling, consider the solid-related subsystem

\[
\bm{x}_{s,k}=
\begin{bmatrix}
\mathrm{SOC}_k & \Delta x_{s,p,k} & \Delta x_{s,n,k}
\end{bmatrix}^{T}.
\]

The corresponding state-transition eigenvalues are
\begin{equation}
a_1=1,\qquad
a_p=1-\frac{\Delta t}{\tau_{s,p}},\qquad
a_n=1-\frac{\Delta t}{\tau_{s,n}}.
\end{equation}

Let the local output sensitivity to this subsystem be

\[
\bm{C}_s =
\begin{bmatrix}
c_{\mathrm{SOC}} & c_p & c_n
\end{bmatrix}.
\]

If $\bm{C}_s$ is locally approximated as constant over a short horizon, the three-step observability submatrix is

\begin{equation}
\bm{\mathcal{O}}_{s}
=
\begin{bmatrix}
c_{\mathrm{SOC}} & c_p & c_n\\
c_{\mathrm{SOC}} & c_p a_p & c_n a_n\\
c_{\mathrm{SOC}} & c_p a_p^2 & c_n a_n^2
\end{bmatrix}.
\end{equation}

Its determinant is

\begin{equation}
\det(\bm{\mathcal{O}}_{s})
=
c_{\mathrm{SOC}}c_pc_n
(a_p-1)(a_n-1)(a_n-a_p).
\label{eq:solid_observability_det}
\end{equation}

Therefore, the solid-related subsystem is locally observable if
\begin{equation}
c_{\mathrm{SOC}}\neq0,\qquad c_p\neq0,\qquad c_n\neq0,\qquad
a_p\neq1,\qquad a_n\neq1,\qquad a_p\neq a_n.
\end{equation}

For the NMC/graphite cell considered in this study, the OCV map used in the IESP model has nonzero local slopes over the operating SOC range. Thus, the terminal voltage remains sensitive to SOC and electrode stoichiometric variations. In addition, the identified effective time constants \(\tau_{s,p}\) and \(\tau_{s,n}\) are different, so \(a_p\neq a_n\). Hence, the coupling of positive and negative electrode concentration-deviation states in the terminal-voltage equation does not structurally cause a rank loss of the solid-related observability submatrix.

For the electrolyte concentration-variation states \(\Delta c_1\) and \(\Delta c_2\), both are governed by the same electrolyte time constant \(\tau_e\). Therefore, under a constant-output linearization at a fixed operating point, their individual observability may be weak. However, in the proposed observer, these states are used as reduced-order polarization states for voltage correction rather than as independently interpreted physical concentration fields. Moreover, under dynamic current profiles such as UDDS and NEDC, the nonlinear concentration-polarization term
\begin{equation}
\eta_{\mathrm{con}}(t)
=
\frac{2RT}{F}(1-t_+)
\ln\left(
\frac{c_0+\Delta c_1(t)}
{c_0-\Delta c_2(t)}
\right)
\end{equation}
induces trajectory-dependent sensitivities:
\begin{equation}
\frac{\partial h}{\partial \Delta c_1}
=
-\frac{2RT}{F}(1-t_+)
\frac{1}{c_0+\Delta c_1},
\qquad
\frac{\partial h}{\partial \Delta c_2}
=
-\frac{2RT}{F}(1-t_+)
\frac{1}{c_0-\Delta c_2}.
\label{eq:electrolyte_sensitivity}
\end{equation}

Thus, the relevant observability condition is the finite-horizon trajectory-dependent condition in Eq.~\eqref{eq:app_observability_rank}, rather than a constant single-point linearization. This is consistent with the SOC-oriented purpose of the proposed reduced-order observer.

For the online-estimated parameter vector

\begin{equation}
\bm{\theta}_k =
\begin{bmatrix}
D_p,\;D_n,\;Q_{\mathrm{all}},\;x_{s,p,0},\;x_{s,n,0}
\end{bmatrix}^{T},
\end{equation}

local practical identifiability can be assessed through the terminal-voltage sensitivity matrix over a data window:

\begin{equation}
\bm{\mathcal{S}}_{\theta}
=
\begin{bmatrix}
\frac{\partial h(t_1)}{\partial \theta_1} &
\frac{\partial h(t_1)}{\partial \theta_2} &
\cdots &
\frac{\partial h(t_1)}{\partial \theta_{n_\theta}}\\
\frac{\partial h(t_2)}{\partial \theta_1} &
\frac{\partial h(t_2)}{\partial \theta_2} &
\cdots &
\frac{\partial h(t_2)}{\partial \theta_{n_\theta}}\\
\vdots & \vdots & \ddots & \vdots\\
\frac{\partial h(t_N)}{\partial \theta_1} &
\frac{\partial h(t_N)}{\partial \theta_2} &
\cdots &
\frac{\partial h(t_N)}{\partial \theta_{n_\theta}}
\end{bmatrix}.
\label{eq:param_sensitivity_matrix}
\end{equation}

A necessary local identifiability condition for the selected parameter subset is

\begin{equation}
\mathrm{rank}\left(\bm{\mathcal{S}}_{\theta}\right)=n_{\theta}.
\label{eq:param_rank_condition}
\end{equation}

Equivalently, the Fisher-information-type matrix

\begin{equation}
\bm{\mathcal{F}}_{\theta}=\bm{\mathcal{S}}_{\theta}^{T}\bm{\mathcal{S}}_{\theta}
\end{equation}

should be nonsingular.

The parameter subset used in this work is selected to improve practical identifiability. As shown by the sensitivity analysis in Appendix~\ref{app:param_sensitivity}, \(D_p\), \(D_n\), \(Q_{\mathrm{all}}\), \(x_{s,p,0}\), and \(x_{s,n,0}\) have the strongest influence on the terminal-voltage response under the UDDS condition. These parameters affect different physical aspects of the reduced-order electrochemical model: \(D_p\) and \(D_n\) determine the effective electrode stoichiometric ranges, \(Q_{\mathrm{all}}\) controls the capacity scaling in the state dynamics, and \(x_{s,p,0}\) and \(x_{s,n,0}\) determine the initial OCV level. Estimating this compact subset, instead of all IESP parameters, reduces parameter collinearity and improves numerical stability.

\clearpage
\setcounter{equation}{0}
\setcounter{figure}{0}
\setcounter{table}{0}

\refstepcounter{section}
\section*{Appendix \thesection. Parameter Sensitivity Analysis under the UDDS Condition}
\label{app:param_sensitivity}
\addcontentsline{toc}{section}{Appendix \thesection. Parameter Sensitivity Analysis under the UDDS Condition}

A one-at-a-time local parameter sensitivity analysis was performed under the UDDS condition to justify the selection of the online-estimated parameter subset in the proposed DSMO. Since the terminal-voltage residual is used as the feedback signal for both state correction and parameter adaptation, the analysis focuses on the influence of each IESP model parameter on the terminal-voltage response.

First, the IESP model was simulated under the UDDS current profile using the nominal parameter set. The resulting terminal-voltage trajectory was taken as the baseline voltage \(U_{\mathrm{base}}(t_k)\). Then, each parameter was perturbed individually within \(\pm20\%\) of its nominal value, while all other parameters were kept unchanged. For each parameter, \(N_p=100\) evenly spaced values were tested in the interval \([0.8p_j^{0},\,1.2p_j^{0}]\), where \(p_j^{0}\) denotes the nominal value of the \(j\)-th parameter. Since \(N_{\theta}=14\) parameters were analyzed, the total number of model simulations was
\begin{equation}
N_{\mathrm{sim}} = 1 + N_{\theta}N_p = 1 + 14 \times 100 = 1401,
\end{equation}
including one baseline simulation and 1400 one-at-a-time perturbed simulations.

\begin{figure}[H]
\centering
\includegraphics[width=\linewidth]{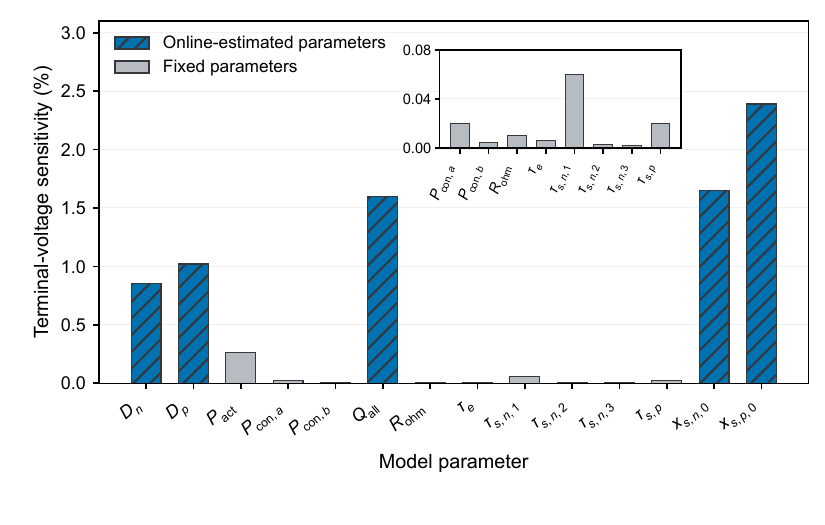}
\caption{Local terminal-voltage sensitivity under UDDS. Blue hatched bars identify the five online-estimated parameters; gray bars identify fixed parameters. The inset resolves weak nonzero sensitivities.}
\label{fig:param_sensitivity_udds}
\end{figure}

For each parameter \(p_j\), the terminal-voltage sensitivity was calculated as
\begin{equation}
S_{U,j}
=
\frac{100}{U_{\mathrm{cell}}}
\sqrt{
\frac{1}{N_p N_t}
\sum_{r=1}^{N_p}
\sum_{k=1}^{N_t}
\left[
U_{\mathrm{base}}(t_k)-U_{j}^{(r)}(t_k)
\right]^2
},
\label{eq:sensitivity_voltage}
\end{equation}
where \(U_{j}^{(r)}(t_k)\) is the terminal-voltage trajectory obtained when the \(j\)-th parameter takes its \(r\)-th perturbed value, \(N_t\) is the number of sampling points in the UDDS profile, and \(U_{\mathrm{cell}}=4.2~\mathrm{V}\) is used as the voltage normalization scale. The sensitivity $S_{U,j}$ is reported in percent, consistently with Figure~\ref{fig:param_sensitivity_udds}. A larger \(S_{U,j}\) indicates that the corresponding parameter has a stronger influence on the terminal-voltage output.

As shown in Figure~\ref{fig:param_sensitivity_udds}, the parameters \(D_n\), \(D_p\), \(Q_{\mathrm{all}}\), \(x_{s,n,0}\), and \(x_{s,p,0}\) exhibit the highest terminal-voltage sensitivities under the UDDS condition. These parameters directly affect the electrode stoichiometry evolution, the effective capacity, and the initial OCV level, and therefore have a dominant influence on both the voltage response and SOC estimation. In contrast, \(P_{\mathrm{act}}\), \(P_{\mathrm{con},a}\), \(P_{\mathrm{con},b}\), \(R_{\mathrm{ohm}}\), \(\tau_e\), \(\tau_{s,n,1}\), \(\tau_{s,n,2}\), \(\tau_{s,n,3}\), and \(\tau_{s,p}\) show relatively lower terminal-voltage sensitivity in this local analysis.

Based on this result, the online parameter estimator is restricted to the following dominant parameter subset:

\begin{equation}
\bm{\theta}_k =
\begin{bmatrix}
D_p,\; D_n,\; Q_{\mathrm{all}},\; x_{s,p,0},\; x_{s,n,0}
\end{bmatrix}^{T}.
\label{eq:selected_param_subset_supp}
\end{equation}

This selection reduces the dimension of the parameter estimator, avoids unnecessary adaptation of weakly sensitive parameters, and improves numerical stability for online implementation. It should be noted that this analysis is a sensitivity analysis around the identified nominal parameter set, rather than a global sensitivity analysis. Therefore, the result is used here to support the practical selection of online-estimated parameters within the operating and parameter ranges considered in this study.

\clearpage
\setcounter{equation}{0}
\setcounter{figure}{0}
\setcounter{table}{0}

\refstepcounter{section}
\section*{Appendix \thesection. Stability Analysis of the ADZ-DSMO}
\label{app:stability_analysis}
\addcontentsline{toc}{section}{Appendix \thesection. Stability Analysis of the ADZ-DSMO}

This appendix provides the stability analysis related to the proposed ADZ-DSMO. The analysis is organized into three parts. First, the practical stability of the residual-driven state SMO is discussed under bounded uncertainty and measurement noise. Second, the Lyapunov-guided adaptive dead-zone boundary is obtained. Third, the bounded conditional update property of the gated parameter SMO is clarified.

\subsection*{C.1. Practical Stability of the Residual-Driven State SMO}
\label{app:state_smo_stability}

The state SMO used in this work is written in the one-step form as

\begin{equation}
\hat{\bm{x}}_{k+1}
=
\bm{A}\hat{\bm{x}}_{k}
+\bm{B} i_{k}
+\bm{K}\,\operatorname{sgn}(e_{y,k})
+\bm{L} e_{y,k}.
\label{eq:app_state_smo_stability_update}
\end{equation}

The corresponding plant dynamics are

\begin{equation}
\bm{x}_{k+1}
=
\bm{A} \bm{x}_k
+
\bm{B} i_k
+
\bm{w}_k,
\label{eq:app_plant_dynamics}
\end{equation}

where $\bm{w}_k$ represents bounded model uncertainty and unmodeled dynamics.

Define the state-estimation error as

\begin{equation}
\tilde{\bm{x}}_k=\bm{x}_k-\hat{\bm{x}}_k .
\end{equation}

Combining Eqs.~\eqref{eq:app_state_smo_stability_update} and \eqref{eq:app_plant_dynamics} gives

\begin{equation}
\tilde{\bm{x}}_{k+1}
=
\bm{A}\tilde{\bm{x}}_{k}
-
\bm{K}\operatorname{sgn}(e_{y,k})
-
\bm{L} e_{y,k}
+
\bm{w}_k .
\label{eq:app_state_error_raw}
\end{equation}

Under a local linearization of the nonlinear terminal-voltage map, the measurable voltage residual can be approximated as

\begin{equation}
e_{y,k}
=
\bm{C}\tilde{\bm{x}}_{k}
+
v_k
+
\xi_k,
\label{eq:app_residual_linear}
\end{equation}

where $\bm{C}$ is the nominal output Jacobian, \(v_k\) is bounded measurement noise, and \(\xi_k\) denotes the bounded local linearization error and residual model mismatch.

Substituting Eq.~\eqref{eq:app_residual_linear} into Eq.~\eqref{eq:app_state_error_raw} yields

\begin{equation}
\tilde{\bm{x}}_{k+1}
=
\bm{A}_c\tilde{\bm{x}}_{k}
+
\bm{d}_k,
\qquad
\bm{A}_c=\bm{A}-\bm{L}\bm{C},
\label{eq:app_state_error_ac}
\end{equation}

where

\begin{equation}
\bm{d}_k
=
\bm{w}_k
-
\bm{K}\operatorname{sgn}(e_{y,k})
-
\bm{L}(v_k+\xi_k).
\label{eq:app_dk_def}
\end{equation}

Assume that $\bm{w}_k$, \(v_k\), and \(\xi_k\) are bounded, i.e.,

\[
\|\bm{w}_k\|\le \bar{w},\qquad |v_k|\le \bar{v},\qquad |\xi_k|\le \bar{\xi}.
\]

Since \(\operatorname{sgn}(e_{y,k})\) is bounded, $\bm{d}_k$ is also bounded:

\begin{equation}
\|\bm{d}_k\|
\le
\bar{w}
+
\|\bm{K}\|
+
\|\bm{L}\|(\bar{v}+\bar{\xi})
\triangleq
\bar{d}.
\label{eq:app_d_bound}
\end{equation}

If the observer gain $\bm{L}$ is selected such that $\bm{A}_c$ is Schur stable, then for any $\bm{Q}_s\succ0$, there exists a symmetric positive-definite matrix $\bm{P}_s\succ0$ satisfying the discrete Lyapunov equation

\begin{equation}
\bm{A}_c^{\top}\bm{P}_s\bm{A}_c-\bm{P}_s=-\bm{Q}_s .
\label{eq:app_state_lyap_eq}
\end{equation}

Choose the Lyapunov function

\begin{equation}
V_k=\tilde{\bm{x}}_{k}^{\top}\bm{P}_s\tilde{\bm{x}}_{k}.
\label{eq:app_state_lyap_function}
\end{equation}

Using Eq.~\eqref{eq:app_state_error_ac}, the Lyapunov difference satisfies

\begin{align}
\Delta V_k
&=
V_{k+1}-V_k \nonumber\\
&=
\tilde{\bm{x}}_{k}^{\top}
(\bm{A}_c^{\top}\bm{P}_s\bm{A}_c-\bm{P}_s)
\tilde{\bm{x}}_{k}
+
2\tilde{\bm{x}}_{k}^{\top}\bm{A}_c^{\top}\bm{P}_s \bm{d}_k
+
\bm{d}_k^{\top}\bm{P}_s \bm{d}_k \nonumber\\
&\le
-\lambda_{\min}(\bm{Q}_s)\|\tilde{\bm{x}}_{k}\|^2
+
2\|\bm{A}_c^{\top}\bm{P}_s\|\|\tilde{\bm{x}}_{k}\|\bar{d}
+
\|\bm{P}_s\|\bar{d}^{2}.
\label{eq:app_state_dv_bound}
\end{align}

Therefore, \(\Delta V_k<0\) when $\|\tilde{\bm{x}}_{k}\|$ is sufficiently large. This implies that the state-estimation error is ultimately bounded under bounded model uncertainty, measurement noise, and local linearization error. In the nominal noise-free case, the ultimate bound decreases accordingly. This establishes the practical stability of the residual-driven discrete-time state SMO used in this work.

\subsection*{C.2. Lyapunov Derivation of the Adaptive Dead-Zone Boundary}
\label{app:lyapunov_deadzone_proof}

This subsection derives a forcing-aware one-step Lyapunov bound and the nominal coefficient ratio used in the adaptive boundary. The observer matrices and gains are held fixed in this derivation, and $\bm{A}$ is assumed nonsingular. Define

\begin{equation}
\begin{aligned}
\hat{\bm{z}}_k
&=\hat{\bm{x}}_k+\bm{r}_k,\\
\bm{r}_k
&=\bm{A}^{-1}(\bm{B}i_k+\bm{K}\phi_k),
\qquad \phi_k=\operatorname{sgn}(e_{y,k}).
\end{aligned}
\label{eq:app_z_def}
\end{equation}

The local voltage linearization used to select a nominal Jacobian is

\begin{equation}
\begin{aligned}
e_{y,k}&\approx\bm{C}_k\tilde{\bm{x}}_k+v_k,\\
\bm{C}_k&=
\left.\frac{\partial h(\bm{x},i_k,\bm{\theta}_k)}{\partial\bm{x}}
\right|_{\bm{x}=\hat{\bm{x}}_k},
\qquad \tilde{\bm{x}}_k=\bm{x}_k-\hat{\bm{x}}_k.
\end{aligned}
\label{eq:app_linearization}
\end{equation}

A fixed nominal Jacobian $\bm{C}$ defines the matrix

\begin{equation}
\bm{A}_c=\bm{A}-\bm{L}\bm{C}.
\label{eq:app_Ac}
\end{equation}

Substitution of Eq.~\eqref{eq:app_z_def} into the actual observer update gives $\hat{\bm{x}}_{k+1}=\bm{A}\hat{\bm{z}}_k+\bm{L}e_{y,k}$. Applying the same auxiliary-variable definition at the next step therefore yields

\begin{equation}
\begin{aligned}
\hat{\bm{z}}_{k+1}
&=\bm{A}\hat{\bm{z}}_k+\bm{L}e_{y,k}+\bm{r}_{k+1},\\
\bm{r}_{k+1}
&=\bm{A}^{-1}(\bm{B}i_{k+1}+\bm{K}\phi_{k+1}).
\end{aligned}
\label{eq:app_z_dynamics}
\end{equation}

Consequently, the exact equivalent form is

\[
\hat{\bm{z}}_{k+1}
=\bm{A}_c\hat{\bm{z}}_k+\bm{u}_k,
\qquad
\bm{u}_k=\bm{L}e_{y,k}+\bm{s}_k,
\]
\[
\bm{s}_k=\bm{L}\bm{C}\hat{\bm{z}}_k+\bm{r}_{k+1}.
\]

The term $\bm{s}_k$ retains the state-dependent contribution and the next-step shift; neither is set to zero in the following bound. Choose

\begin{equation}
V_k=\hat{\bm{z}}_k^{\top}\bm{P}\hat{\bm{z}}_k,
\qquad \bm{P}\succ0.
\label{eq:app_V}
\end{equation}

The construction assumes a Schur-stable $\bm{A}_c$ and a symmetric positive-definite solution of

\begin{equation}
\bm{A}_c^{\top}\bm{P}\bm{A}_c-\bm{P}=-\bm{Q},
\qquad\bm{Q}\succ0.
\label{eq:app_lyap_eq}
\end{equation}

The Lyapunov difference is then exactly

\begin{equation}
\begin{aligned}
\Delta V_{k+1}
&=V_{k+1}-V_k\\
&=-\hat{\bm{z}}_k^{\top}\bm{Q}\hat{\bm{z}}_k
+2\hat{\bm{z}}_k^{\top}\bm{A}_c^{\top}\bm{P}\bm{u}_k
+\bm{u}_k^{\top}\bm{P}\bm{u}_k.
\end{aligned}
\label{eq:app_dV_expand}
\end{equation}

For any $\beta>0$, weighted Young's inequality gives

\begin{equation}
\begin{aligned}
2\hat{\bm{z}}_k^{\top}\bm{A}_c^{\top}\bm{P}\bm{u}_k
\le{}&\beta\hat{\bm{z}}_k^{\top}
\bm{A}_c^{\top}\bm{P}\bm{A}_c\hat{\bm{z}}_k\\
&+\beta^{-1}\bm{u}_k^{\top}\bm{P}\bm{u}_k.
\end{aligned}
\label{eq:app_young}
\end{equation}

This follows by applying $2\bm{a}^{\top}\bm{b}\le\beta\|\bm{a}\|^2+\beta^{-1}\|\bm{b}\|^2$ to $\bm{a}=\bm{P}^{1/2}\bm{A}_c\hat{\bm{z}}_k$ and $\bm{b}=\bm{P}^{1/2}\bm{u}_k$. Substituting Eq.~\eqref{eq:app_lyap_eq} yields

\begin{equation}
\begin{aligned}
\Delta V_{k+1}
&\le-\hat{\bm{z}}_k^{\top}\bm{H}_{\beta}\hat{\bm{z}}_k
+(1+\beta^{-1})\bm{u}_k^{\top}\bm{P}\bm{u}_k,\\
\bm{H}_{\beta}&=(1+\beta)\bm{Q}-\beta\bm{P}.
\end{aligned}
\label{eq:app_dV_bound}
\end{equation}

Select $\beta$ so that $\bm{H}_{\beta}\succ0$. For $\bm{L}\ne\bm{0}$, define

\begin{equation}
\begin{aligned}
c_z&=\lambda_{\min}(\bm{H}_{\beta})>0,\\
c_e&=(1+\beta^{-1})\bm{L}^{\top}\bm{P}\bm{L}>0.
\end{aligned}
\label{eq:app_cz_ce}
\end{equation}

The triangle inequality in the $\bm{P}$-weighted norm gives

\[
\sqrt{\bm{u}_k^{\top}\bm{P}\bm{u}_k}
\le
\sqrt{\bm{L}^{\top}\bm{P}\bm{L}}\,|e_{y,k}|
+\sqrt{\bm{s}_k^{\top}\bm{P}\bm{s}_k}.
\]

Let $\sigma_k=\sqrt{\bm{s}_k^{\top}\bm{P}\bm{s}_k/(\bm{L}^{\top}\bm{P}\bm{L})}$. Consequently,

\[
\Delta V_{k+1}
\le-c_z\|\hat{\bm{z}}_k\|^2
+c_e(|e_{y,k}|+\sigma_k)^2.
\]

A sufficient condition for strict one-step decrease is therefore

\begin{equation}
|e_{y,k}|+\sigma_k
<\sqrt{\frac{c_z}{c_e}}\,\|\hat{\bm{z}}_k\|.
\label{eq:app_ey_bound}
\end{equation}

When $\bm{s}_k=\bm{0}$, Eq.~\eqref{eq:app_ey_bound} reduces to the nominal residual condition. Its coefficient ratio motivates the online boundary

\begin{equation}
\delta_k
=\sqrt{\frac{c_z}{c_e}}
\left\|\hat{\bm{x}}_k+\bm{A}^{-1}\bm{B}i_k
+\bm{A}^{-1}\bm{K}\operatorname{sgn}(e_{y,k})\right\|.
\label{eq:app_delta}
\end{equation}

The parameter-update gate uses the nominal criterion

\begin{equation}
|e_{y,k}|\le\delta_k.
\label{eq:app_gate_condition}
\end{equation}

Equation~\eqref{eq:app_ey_bound} is the forcing-aware decrease certificate, whereas Eq.~\eqref{eq:app_gate_condition} defines the nominal online gating rule. In general $\bm{s}_k\ne\bm{0}$, so the latter does not imply the former. The state-error boundedness result is established separately in Appendix~\ref{app:state_smo_stability}, and the gated parameter-increment bound is given in Appendix~\ref{app:gated_param_smo_boundedness}.

For the identity weighting choice

\begin{equation}
\bm{Q}=\bm{I}_5,
\label{eq:app_Q_choice}
\end{equation}

the positive-definiteness requirement is equivalent to

\begin{equation}
\lambda_{\max}(\bm{P})<1+\beta^{-1}.
\label{eq:app_beta_admissibility}
\end{equation}

The candidate selection

\begin{equation}
\beta=10^{-4}
\label{eq:app_beta_choice}
\end{equation}

is therefore admissible when the actual Lyapunov solution satisfies $\lambda_{\max}(\bm{P})<10001$. If this test fails, a smaller positive $\beta$ must be selected. The matrix $\bm{P}$ must be obtained from the same nominal matrix and discrete sampling interval used in Eq.~\eqref{eq:app_lyap_eq}; positive definiteness of an independently specified weighting matrix is not sufficient.

\subsection*{C.3. Boundedness of the Gated Parameter SMO}
\label{app:gated_param_smo_boundedness}

The parameter SMO is updated conditionally according to the adaptive dead-zone gate. The candidate parameter update is

\begin{equation}
\hat{\bm{\theta}}_{k+1}^{\mathrm{cand}}
=
\hat{\bm{\theta}}_{k}
+
\bm{K}^{\theta}\operatorname{sgn}(e_{y,k})
+
\bm{L}^{\theta}e_{y,k}.
\label{eq:app_param_candidate}
\end{equation}

The implemented parameter update can be written as

\begin{equation}
\hat{\bm{\theta}}_{k+1}
=
\hat{\bm{\theta}}_{k}
+
g_k
\left[
\bm{K}^{\theta}\operatorname{sgn}(e_{y,k})
+
\bm{L}^{\theta}e_{y,k}
\right],
\label{eq:app_param_gated_update}
\end{equation}

where \(g_k\in\{0,1\}\) is the adaptive dead-zone gate:
\begin{equation}
g_k=
\begin{cases}
1, & |e_{y,k}|\le\delta_k,\\
0, & |e_{y,k}|>\delta_k.
\end{cases}
\label{eq:app_gate_signal}
\end{equation}

When \(g_k=0\), the parameter estimate is frozen and no residual-induced parameter correction is introduced:

\begin{equation}
\hat{\bm{\theta}}_{k+1}=\hat{\bm{\theta}}_k .
\end{equation}

When $g_k=1$, the residual satisfies the selected gate condition in Eq.~\eqref{eq:app_gate_condition}. Therefore, the one-step parameter variation is bounded by

\begin{equation}
\begin{aligned}
\left\|
\hat{\bm{\theta}}_{k+1}
-
\hat{\bm{\theta}}_{k}
\right\|
&\le
\left\|\bm{K}^{\theta}\operatorname{sgn}(e_{y,k})\right\|
+
\left\|\bm{L}^{\theta}e_{y,k}\right\| \\
&\le
\left\|\bm{K}^{\theta}\right\|
+
\left\|\bm{L}^{\theta}\right\|\delta_k .
\end{aligned}
\label{eq:app_param_increment_bound}
\end{equation}

From Eq.~\eqref{eq:app_delta}, $\delta_k$ is computed from the estimated state, measured current, observer matrices, and offline-computed design constants. Under bounded current input and bounded estimated states, $\delta_k$ is finite. Hence, the one-step update of the parameter SMO is bounded whenever the gate is active. When the residual exceeds the selected adaptive boundary, the gate is closed and the parameter update is suppressed.

This analysis shows that the proposed adaptive dead-zone gate provides bounded conditional parameter adaptation and prevents large transient voltage residuals from being directly propagated into the parameter estimator. It should be noted that this bounded-update result does not imply unconditional global convergence of the estimated parameters. Exact parameter convergence requires additional persistent-excitation and local identifiability conditions, as discussed in Appendix~\ref{app:observability_identifiability}. The purpose of the gated parameter SMO is therefore to improve practical online adaptation and SOC-estimation robustness while avoiding unreliable parameter corrections.

\bibliographystyle{elsarticle-num} 
\bibliography{references}

\end{document}